\documentclass{article}

\usepackage{vmargin,amssymb,amsmath,epsfig}

\newcommand{\eqsp}{\, = \,}  
\newcommand{\rar}{\, \rightarrow \,}
\newcommand{\beq}{\begin{equation}}
\newcommand{\eeq}{\end{equation}}
\newcommand{\la}{\langle}
\newcommand{\ra}{\rangle}
\newcommand{\ep}{\epsilon}

\begin{document}

\thispagestyle{empty}

\begin{flushright} HU-MATH-2025-01 \end{flushright}

\vskip 4 cm

\begin{center}

{\huge \textbf{The off-shell one- and two-loop box recovered from intersection theory}}

\vskip 1.5 cm

{\large B.~Eden}
\vskip 1 cm

$^b$ Institut f\"ur Mathematik und Physik, Humboldt-Universit\"at zu Berlin, \\ Zum gro{\ss}en Windkanal 2, 12489 Berlin, Germany \\[2 mm]
\vskip 1 cm

e-mail: eden@math.hu-berlin.de

\end{center}

\vskip 3 cm

\textbf{Abstract:}  

We advertise intersection theory for generalised hypergeometric functions as a means of evaluating Mellin-Barnes representations. As an example, we study two-parameter representations of the off-shell one- and two-loop box graphs in exactly four-dimensional configuration space. Closing the integration contours for the MB parameters we transform these into double sums. Polygamma functions in the MB representation of the double box and the occurrence of higher poles are taken into account by parametric differentiation. Summing over any one of the counters results into a $_{p+1}F_p$ that we replace by its Euler integral representation. The process can be repeated a second time and results in a two- or four-parameter Euler integral, respectively. We use intersection theory to derive Pfaffian systems of equations on related sets of master integrals and solve for the box and double box integrals reproducing the known expressions. Finally, we use a trick to re-derive the double box from a two-parameter Euler integral. This second computation requires only very little computing resources.

\newpage

\section{Introduction}

\emph{Gluing contributions} \cite{BKV,shotaThiago1,shotaThiago2} in higher-point \emph{integrability} computations in ${\cal N} \eqsp 4$ super Yang-Mills theory yield sum-integrals akin to partially evaluated Mellin-Barnes (MB) representations (see \cite{theMB} and references therein), by which we mean that some of the integration contours are closed over a half plane and series of residues are picked. Lacking alternative approaches, one will then do the same for the actual integrals creating a multiple sum.

In the examples in \cite{usFivePoint,gluingInter} any individual summmation is of hypergeometric $_{p+1}F_p$ type and may be replaced by its Euler integral representation. Pulling the integration in the Euler parameters through the remaining summations, summands are obtained that are built from the same elements as before, i.e. they contain powers of some polynomials and $\Gamma$ functions. Therefore, the procedure can be iterated. A multi-parameter Euler integral arises which can often be \emph{directly integrated} to generalised polylogarithms. Alternatively \cite{gluingInter}, one may draw upon \emph{intersection theory} for hypergeometric functions \cite{inter1,inter2}.

In this article we analyse the two lowest conformal ladder integrals (aka. the off-shell box and double box) by this summation/integration technique to shed light on the question whether the method applies to Feynman integrals. While direct integration fails, the answer is affirmative if intersection theory is invoked: we are able to reproduce the lowest two cases of the known results for the off-shell ladder diagrams \cite{davydychev1,davydychev2}. 

\begin{center}
\hskip 0.5 cm \includegraphics[width = 9.5 cm]{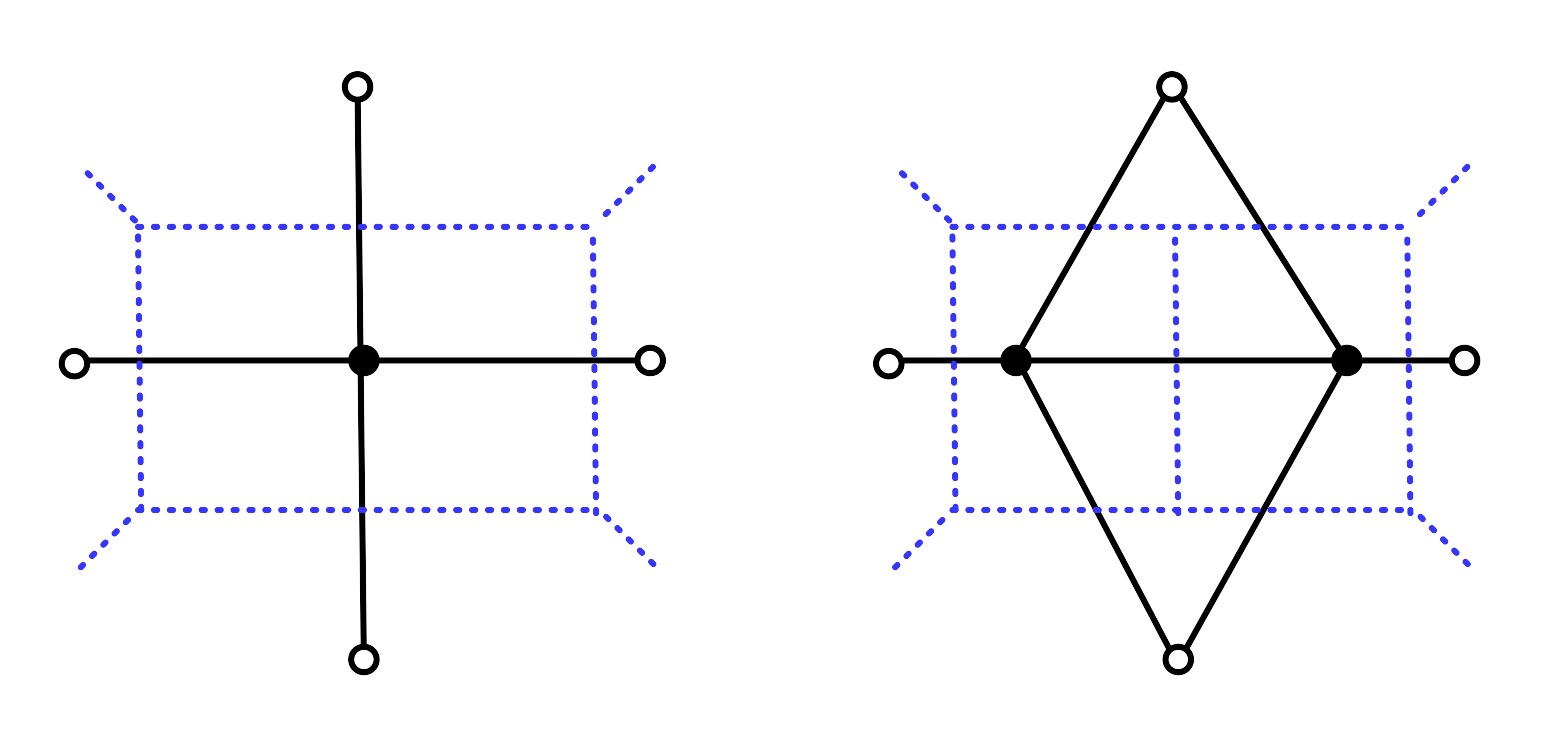} \\
\quad Figure 1: the one and two-loop box in configuration space (black) \\ and dual momentum space (blue)
\end{center}

Let us focus on the configuration space diagrams in Figure 1. Starting from the left-most point in either panel we label the points in a clockwise sense as $\{x_1, \, x_3, \, x_2, \, x_4\}$. Both integrals are finite, and in four dimensions they are also conformal implying that they only depend on the two cross-ratios
\beq
u \eqsp \frac{x_{13}^2 x_{24}^2}{x_{12}^2 x_{34}^2} \, , \qquad v \eqsp \frac{x_{14}^2 x_{23}^2}{x_{12}^2 x_{34}^2}
\eeq
apart from a rational factor needed to arrange for the correct conformal weights. As a consequence, we may send one point to infinity without losing information: for instance, the limit $x_4 \rar \infty$ amounts to the replacement
\beq
u \rar \frac{x_{13}^2}{x_{12}^2} \, , \qquad v \rar \frac{x_{23}^2}{x_{12}^2}
\eeq 
which does not reduce the number of independent variables. In the graphs, all propagators attached to the lower point are amputated, demonstrating the equivalence \cite{davydychev1,davydychev2} of the left-over vertex diagrams with the former box functions.

In the following we shall use the parameterisation 
\beq
u \eqsp z \bar z \, , \qquad v \eqsp (1-z)(1-\bar z) \qquad \Rightarrow \qquad \sqrt{(1-u+v)^2 - 4 v} \eqsp z - \bar z \label{defBZ}
\eeq
commonly employed to linearise the root function in the right equation, which occurs in the original writing of the results \cite{davydychev1,davydychev2}. In the new variables, the two vertex functions are
\beq
I_1 \eqsp \frac{1}{x_{12}^2 (z - \bar z)} \bigl( 2 \, \text{Li}_2[z] - 2 \, \text{Li}_2[\bar z]  + \log[z \bar z] \, ( \log[1-z] - \log[1-\bar z] ) \bigr) \, , \label{davyBox1}
\eeq
\beq
I_2 \eqsp - \frac{1}{x_{12}^2 \, (z - \bar z)} \Phi^{(2)}\left[\frac{z}{z-1}, \, \frac{\bar z}{\bar z - 1} \right]  \, , \quad \Phi^{(2)}[z,\bar z] \eqsp \sum_{r \eqsp 0}^{2} \frac{(-1)^r (4-r)!}{r! \, (2-r)! \, 2!} \, \log^r[z \bar z] \, (\mathrm{Li}_{4-r}[z] - \mathrm{Li}_{4-r}[\bar z]) \, . \label{ExactDB}
\eeq
Remarkably, all higher off-shell ladder diagrams can be recursively determined \cite{davydychev2} and are given by similar expressions. We do not attempt to reproduce these results in the present work.

\newpage

The article is organised as follows: in Section
\begin{itemize}
\item \ref{oneLoopMBEuler} we make an MB representation for the one-loop box into a double sum over residues and re-sum into a two-parameter Euler integral,
\item \ref{introIntersect} we review bivariate intersection theory and apply it to the one-loop box,
\item \ref{doubleIntersectLong} we derive two four-parameter Euler-integrals for the parts of the MB representation \eqref{MBrepI2} of the double box, construct Pfaffian equations and solve these in terms of Goncharov logarithms. We reproduce the symbol \cite{symbolGonchiLog1,symbolGonchiLog2} of \eqref{ExactDB},
\item \ref{doubleIntersectShort} we use a parameter shift to substantially simplify the latter computation falling upon the same functions and symbols.
\end{itemize}
Finally, we present some conclusions. Appendix A reviews how to build MB representations. Formula \eqref{MBrepI2} is derived there. Appendix B collects the intersection matrices, higher connections, and Pfaffian matrices for the shorter computation of Section \ref{doubleIntersectShort}. Bulkier expressions are relegated to the ancillary material to this publication, i.e. the same items for the longer computation of Section \ref{doubleIntersectLong}, and the functions and symbols for the two parts (which agree between the two versions of the calculation) composing that of the double box.

\section{Euler integrals for the off-shell one-loop box} \label{oneLoopMBEuler}

The Feynman diagram has the simple and appealing MB representation \cite{davydychev1}
\beq
I_1 \eqsp \frac{1}{(2 \pi i)^2 \, x_{12}^2} \int dz_1 dz_2 \, u^{z_1} v^{z_2} \, \Gamma[-z_1]^2 \Gamma[-z_2]^2 \Gamma[1+z_1+z_2] ^2 \, . \label{MBrepI1}
\eeq
The integration domains for both MB parameters are straight lines parallel to the imaginary axis. With the aim of stating the basic steps of the approach and the relevant formulae we include a derivation of \eqref{MBrepI1} in Appendix A. 

One can close both integration contours to the right, or one to the right and one to the left, leading to three distinct series expansions of $I_1$ in $\{u,v\}, \, \{u/v,1/v\}, \, \{1/u,v/u\}$, respectively. Closing both contours to the right we pick residues from $\Gamma[-z_1]^2 \Gamma[-z_2]^2$, so $z_1, z_2 \rar n_1, n_2$ for integer $n_i$. The position of the real part of the  original straight line integrations determines the starting values for the sums over $n_1, \, n_2$. The package  {\tt MBresolve.m} \cite{MBresm} suggests\footnote{These conclusions are not hard to reach for $I_1$. We introduce {\tt MBresolve.m} here because it will become relevant to the application of Barnes' Lemmata for the simplification of the MB representation of $I_2$, see Appendix A.}
\beq
\{ \Re(z_1) \eqsp -0.476559 \, , \Re(z_2) \eqsp -0.0434094 \}  \, . \label{MBRvalI1} 
\eeq
These negative real parts imply that zero is included, thus $n_1, \, n_2 \, \in \, \mathbb{N}_0$.
Given that
\beq
\mathrm{res}_{z = - n} \Gamma[z] \eqsp \frac{(-1)^n}{\Gamma[1+n]}
\eeq
closing both contours to the right we find the expansion
\beq
I_1 \eqsp \frac{1}{x_{12}^2} \sum_{n_1,n_2 \eqsp 0}^\infty \partial_\gamma \partial_\delta \, u^{n_1+\gamma} v^{n_2+\delta} \, \frac{\Gamma[1+n_1+n_2 +\gamma + \delta]^2}{\Gamma[1+n_1+\gamma]^2 \Gamma[1+n_2+\delta]^2} \, ,\qquad \gamma \eqsp 0 \eqsp \delta \label{sumI1}
\eeq 
The differentiation in the parameters $\gamma, \delta$ was introduced because both poles are double.

Note that the undifferentiated sum on the rhs. of the last equation at $\gamma \eqsp 0 \eqsp \delta$ is the double Taylor expansion of $1/\sqrt{(1-u+v)^2 - 4 v}$ with the root function from \eqref{defBZ}. The series converges for small enough $u,v$ as one may numerically check on truncations. It is simple to re-sum at $u \eqsp v$ which suggests the estimate $\sqrt{u^2+v^2} < 1/\sqrt{8}$; the exact region of convergence is larger but unknown to us. Importantly, since a non-vanishing domain of convergence exists we will freely swap differentiation over summation/integration in the following. This assumption is, of course, quite non-trivial and will be justified only a posteriori by the success of our exercise.

The $n_1$ and $n_2$ sums in \eqref{sumI1} both individually yield hypergeometric $_3F_2$ functions, for instance
\beq
\sum_{n_1 \eqsp 0}^\infty u^{n_1} \, \frac{\Gamma[1+n_1+n_2 +\gamma + \delta]^2}{\Gamma[1+n_1+\gamma]^2} \eqsp \frac{\Gamma[1+n_2+\gamma+\delta]^2}{\Gamma[1+\gamma]^2} \, {_3F_2[u]}^{\{1, 1 + n_2 + \gamma + \delta, 1+ n_2 + \gamma + \delta\}}_{\{1 + \gamma, 1 + \gamma\}} \, .
\eeq
We now use the integral representation
\begin{eqnarray}
_3F_2[u]^{\{a_1,a_2,a_3\}}_{\{b_1,b_2\}} &=& \frac{\Gamma[b_1] \Gamma[b_2]}{\Gamma[a_2] \Gamma[a_3] \Gamma[b_1-a_2] \Gamma[b_2-a_3]} \, * \nonumber \\
&& \quad \int_0^1 ds \, dt \, s^{a_2-1}  (1-s)^{b_1-a_2-1} t^{a_3-1} (1-t)^{b_2-a_3-1} (a - s \,  t \,  u)^{-a_1}
\end{eqnarray}
and
\beq
\frac{1}{\Gamma[-n_2-\delta] \Gamma[1+n_2+\delta]} \eqsp \frac{\sin[\pi (n_2 + \delta)]}{\pi} \eqsp \frac{(-1)^{n_2} \sin[\pi \delta]}{\pi} \label{doubleGamma}
\eeq
to see that the $n_2$ sum collapses to a geometric series in the argument $s \, t \, v /((1-s)(1-t))$. Hence
\beq
I_1 \eqsp \partial_\gamma \partial_\delta \frac{\sin[\pi \delta] ^2\, u^\gamma v^\delta}{x_{12}^2}  \int_0^1 \frac{ds \, dt \, s^{\gamma + \delta} (1-s)^{-\delta} t^{\gamma+\delta} (1-t)^{-\delta}}{(1 - s \, t  \, u)(1 - s - t + s \, t - s \, t  \, v)} \, , \qquad \gamma \eqsp 0 \eqsp \delta \, . \label{EulerI1}
\eeq

Two comments are in order. First, ${_{p+1}F_p}^{\{a_i\}}_{\{b_j\}}$ is symmetric w.r.t. permutations of both index sets. In the case at hand this means that up to variable transformations one other Euler-integral can be derived. In that case, the first sum yields $_3F_2$ as before, but the second sum is not of geometric type. It is a $_2F_1$ so that in total a three-parameter integral is obtained. A further two equivalent representations are obtained taking the sum in $n_2$ before that in $n_1$. We will not study the latter three Euler integrals here.

Second, how can we hope to extract a $\gamma^1 \, \delta^1$ component from the Laurent expansion of \eqref{EulerI1}? Simply expanding the numerator in $\gamma, \delta$ we obtain logarithms and thus integrable singularities, so that the entire expression seems of minimal order $\delta^2$ (before differentiation) due to the factor $\sin[\pi \delta]^2$ up front. Thus \emph{direct integration} of the expanded integrand clearly fails. On the other hand, Euler integrals are the true realm of \emph{intersection theory} for hypergeometric functions \cite{inter1}. An application has been given also in the context of ${\cal N} \eqsp 4$ gluings in \cite{gluingInter}, although primarily with the intention of simplifying the integration. However, here the use of intersection theory seems mandatory as it resolves the conundrum about the leading orders of the Laurent expansion revealing a simple pole in $\delta$ of the integral in \eqref{EulerI1}.

\section{Bivariate intersection theory for the one-loop box} \label{introIntersect}

We review the necessary minimum of information closely following \cite{inter3a,inter3b,inter3}, though specialising to the case at hand. Consider 2-forms
\beq
\phi_L \eqsp \hat \phi_L(s,t) \, ds \wedge dt \, , \qquad \phi_R \eqsp \hat \phi_R(s,t) \, ds \wedge dt \, .
\eeq
These forms will be rational and so have a polynomial numerator and a number of denominator polynomials. The aim is to compute integrals of the type
\beq
I \eqsp \iint_0^1 w(s,t) \, \phi_L
\eeq
forming a family, i.e. various integrals which may have some polynomial numerator, and in which the elements of the set of all denominators $\{p_1 \ldots p_n\}$ can be raised to non-negative integer powers. The factor $w$ defining the \emph{twisted cohomology} is the product of all possible denominators raised to some small non-integer power.

To be concrete, in our example \eqref{EulerI1} the possible denominators are
\beq
\{p_1 \ldots p_6\} \eqsp \{ s, \, 1-s, \, t, \, 1-t, \, 1 - s \, t \, u , \, 1 - s - t + s \, t - s \, t \, v \}
\eeq 
and the twist factor is
\beq
w \eqsp s^{\gamma + \delta} (1-s)^{-\delta} t^{\gamma+\delta} (1-t)^{-\delta} p_5^\rho \, p_6^\rho \, .
\eeq
Later, we will rescale $\gamma \rar \gamma \, \ep, \, \delta \rar \delta \, \ep, \, \rho \rar \rho \, \ep$, where $\ep$ plays the r\^ole of a dimensional regulator. The first four factors in $w$ do in fact regularise the integral at the boundaries of the integration region. We may choose the powers of $p_1 \ldots p_4$ as in \eqref{EulerI1}. Yet, similar factors involving $p_5, p_6$ are needed at intermediate stages of the computation. At the end \eqref{EulerI1} is recovered in the limit $\rho \rar 0$.

Intersection theory allows us to decompose the most general integral in the family in terms of a set of masters. Let
\beq
\hat \omega_s \eqsp \partial_s \, \log(w) \, , \qquad \hat \omega_t \eqsp \partial_t \, \log(w) \, . \label{countMs}
\eeq
Then the number of solutions to the simultaneous conditions $\hat \omega_s \eqsp 0 \eqsp \hat \omega_t$ indicates the number of master integrals \cite{howManyMasters}. In our example there are seven.

The decomposition will require two steps: we first single out one of the integrations to start with, say, in the variable $s$. Regarding any form $\phi_L$ as some coefficient only depending on $t$ and a one-form in $s$, we should be able to express it in terms of a basis of master one-forms w.r.t. $s$. Here the relevant number of masters is given by the number of zeroes of the first condition in $\eqref{countMs}$ only. In our application there will be three masters $\{e_1, \, e_2, \, e_3\}$ which are one-forms in $s$. We will choose
\beq
\left\{\frac{ds}{(1-t)(1-s)}, \frac{ds}{p_5}, \, \frac{ds}{p_6} \right\} \, . \label{boxBasS}
\eeq
Note that the second and third masters have simple poles only in both parameters $s,t$. In view of the second integration step in the variable $t$ we have introduced the denominator factor $1-t$ into the first factor. Such \emph{dLog} masters are desirable to limit the computational effort, see below.

If anything works, it must be so that
\beq
\phi_L \eqsp \sum_{i \eqsp 1}^3 \left(\phi_L'(t)_i \, dt \right)  \wedge e_i  \, , \qquad \phi_R \eqsp \sum_{j \eqsp 1}^3 h_j \wedge \left(\phi_R'(t)_j \, dt \right)
\eeq
where $h_i$ are three right masters. It is not necessary to have equal left and right forms but it will be most convenient to identify the two bases \cite{gluingInter}.

The pairing (in our case a $3 \times 3$ matrix) 
\beq
C_s \eqsp \la e_i | h_j \ra
\eeq
of \emph{intersection numbers} is defined as follows: we determine the total set of poles $\{s_1 \ldots s_k\}$ in $s$ (these may depend on outer variables as well as $t$) coming from the collection of denominators. For each pole we replace $s \rar s_m + x$ with a new variable $x$ in which we Taylor expand. The covariant derivative w.r.t. to $s$ is defined by the partial derivative acting under the integral, where the effect of the differentiation on $w$ creates a connection term:
\beq
\nabla_s \eqsp \partial_s + \hat \omega_s \label{nablaS}
\eeq
We next define a \emph{potential} for every $e_i$ in a neighbourhood of a pole $s_m$: let us write $e_i \eqsp \hat e_i \, ds$, so as in the beginning of the section hatting expresses the omission of the differential. Assuming the expansion of $\hat e_i|_{s \eqsp s_m + x}$ to start at a minimal power $min_m-1$ and that we can truncate at some maximum $max_m$ we define the potential as
\beq
\psi_{i,m} \eqsp \sum_{n \eqsp min_m}^{max_m} c_i^{(n)} \, x^n 
\eeq
and solve
\beq
\left(\partial_s + \hat \omega_s|_{s \eqsp s_m + x}\right) \, \psi_{i,m} \eqsp \hat e_i|_{s \eqsp s_m + x} \label{defPsiV}
\eeq
comparing the coefficients of the Taylor expansion of the two sides from $min_m-1$ to $max_m-1$. Finally, the intersection number is:
\beq
C_s \eqsp \sum_{m \eqsp 1}^k res|_{s \eqsp s_m} \, \psi_{i,m} \, \hat h_j|_{s \eqsp s_m + x}
\eeq
It follows that $max_m+1$ is the negative of the minimal order of $\hat h_j$ in the region $m$. It is advisable to choose the masters such that the range of powers in the potentials is minimised. In practice this implies to prefer masters with  simple poles only. We emphasise that the pole at infinity has to be taken account by the rule $s \rar 1/x$ because the problem is projective in nature. These arguments point at dLog masters as the preferred choice.

For a basis of masters, $C_s$ will be invertible. In our example, the inverse $C_s^{-1}$ is nicer than $C_s$ itself:
\beq
C_s^{-1} \eqsp -\frac{1}{\gamma + \delta} \, D_s \left( \begin{array}{ccc} \gamma \delta & \delta \rho & \delta \rho \\ \delta \rho & -(\delta + \gamma + \rho) \rho & -\rho^2 \\ \delta \rho &- \rho^2 & -(\delta + \gamma + \rho) \rho \end{array} \right) D_s \label{boxC1}
\eeq
with the diagonal matrix
\beq
D_s \eqsp \mathrm{diag}\{ 1-t, \, t \, z \bar z, \, 1 - t (z + \bar z - z \bar z) \} \, . \label{boxCoord}
\eeq

The form of $C_s^{-1}$ in \eqref{boxC1} is apparently a hallmark of a well-chosen basis in all computations with Euler integrals: the coordinate dependence \eqref{boxCoord} factors out\footnote{From the point of view of the $s$-integration, $t$ is an outer variable like $z, \bar z$.} while the differentiation/regularisation parameters $\gamma, \, \delta, \, \rho$ define the core part of the intersection numbers. The necessity of introducing the additional regulator $\rho$ becomes apparent because without it the matrix would be singular. In comparison, the computations in \cite{gluingInter} involve only one such parameter, which can then be scaled out as well leaving an array of rational numbers.

Let us return to the abstract discussion. The existence of $C_s^{-1}$ allows us to construct the coefficients $\phi'_L(t)_i$ in a simple way:
\beq
\sum_{j\eqsp1}^3 \la \phi_L | h_j \ra (C_s^{-1})_{ji} \eqsp \phi'_L(t)_i \, dt \, .
\eeq
To unclutter the nomenclature we will refer to the $\phi_L'$ as \emph{$t$-projected left forms}. They are three-component objects. We introduce $t$-projected right masters, too, though without an explicit left multiplication by $C_s^{-1}$:
\beq
\varphi'_R(t)_i \, dt \eqsp \la e_i | \phi_R \ra \, , \qquad \phi_R' \eqsp C_s^{-1} \varphi'_R \, .
\eeq
We further need a $3 \times 3$ second level connection. Let
\beq
\nabla_t \eqsp \partial_t + \hat \omega_t
\eeq
and define
\beq
(\Omega_t)_{ij} \eqsp \sum_{k \eqsp 1}^3 \la \nabla_t e_i | h_k \ra (C_s^{-1})_{kj} \, . \label{makeOmT}
\eeq
where the $s$ intersection number in the angle brackets is constructed just as described for the left forms themselves, i.e. building a potential and then taking residues in its product with the right level 1 masters. In the box example, this connection is cumbersome to write out; we rather include it in the ancillary material. It has poles of maximally first order in $t$ at the points
\beq
t_m \, \in \, \left\{ 1, \, 0, \, \frac{1}{z}, \, \frac{1}{\bar z}, \, \frac{1}{z \bar z}, \, \frac{1}{z + \bar z - z \bar z} \right\}
\eeq
which are the singularities to expand around at level 2. By construction, this covers the set of poles in the $t$-projected left or right quantities. As before, we will replace $t \rar t_m + x$ and expand in the parameter $x$. Also here the pole at infinity cannot be neglected.

For the $t$ intersection numbers we need a potential $\psi$, too, of which a range of powers $min \ldots max$ in $x$ is needed. It must be a three-component object so that we impose a matrix version of equation \eqref{defPsiV}:
\beq
\partial_v \, \psi  + \psi \, . \, \Omega_t|_{t \eqsp t_m + x} \eqsp \hat \phi'_L|_{t \eqsp t_m + x} \label{defPsiW}
\eeq
where $\phi'_L$ on the r.h.s. is also the collection of all three components. Their lowest order $min_m$ determines the start of the expansion of the potential, but to be on the safe side we should take the absolute minimum of the lowest orders  estimates of the three components of $\phi'_L$, which we call $min_m-1$ again since $\Omega_t$ as well as the partial $t$ derivative lower the order in $x$ by 1. On the other hand, the maximum order $max_m$ will be defined by the $t$-projected right master in question, taking the absolute maximum of the three order estimates $-min_m-1$ for the  individual components.

The second level intersection number is
\beq
C_t \eqsp \la \phi_L | \phi_R \ra_t \eqsp \sum_{m \eqsp_1}^l res|_{t \eqsp t_m} \, \psi \, . \, \varphi_R'|_{t \eqsp t_m + x}
\eeq
with $\psi$ defined by \eqref{defPsiW}. Here $\phi_L$ and $\phi_R$ are any member of larger bases of masters completing the level 1 bases of length 3. For true bases, $C_t$ will be invertible, too.

Any two-form can be decomposed in terms of the second level left basis using
\beq
\phi \eqsp \la \phi \, | \, \underline h \, \ra \, C_t^{-1} \, , 
\eeq
where $\underline h$ is the full level 2 right basis. In particular, this is so for derivatives of the left masters w.r.t. outer parameters. Let
\beq
M_y \eqsp \la (\partial_y + \hat \omega_y) \, \underline{e} \, | \, \underline{h} \, \ra \, C_t^{-1} \, , \qquad \hat \omega_y \eqsp \partial_y \, \log(w) \, .
\eeq 
This allows us to write one Pfaffian differential equation
\beq
\partial_y \, \underline{e} \eqsp M_y \, . \, \underline{e} \label{doPfaff}
\eeq
for every outer parameter.

Guessing the most suitable basis is an art. Nonetheless, bilinears of the various denominators do seem natural and turn out to be very useful, see also \cite{gluingInter}. In our off-shell one-loop box example, a near ideal choice is
\beq
 \left\{ \frac{1}{p_5 \, p_6}, \, \frac{1}{(1 - t) p_6}, \,  \frac{1}{(1 - s) p_5}, \, \frac{1}{(1 - s) p_6}, \, \frac{1}{(1 - t) (1 - s)}, \,  \frac{1}{p_5}, \, \frac{1}{p_6}  \right\} ds \, dt \, , \label{boxBasST}
\eeq
where the first form yields the (undifferentiated) box integral \eqref{EulerI1} up to normalisation. The last three entries are as in our first level basis \eqref{boxBasS} although they now carry both differentials. All the masters have simple poles only in $s, t$ apart from the scaling at infinity, which may be of order 0. In particular, with this basis only the two leading orders of the expansion around poles are solicited which substantially reduces the computational load. The decomposition is very quick on a laptop also for the derivatives of the master integrals.

The resulting matrix of second level intersection numbers is
\beq
C_t \eqsp - D_t^{-1} \, C_t' \ D_t^{-1}
\eeq
with the diagonal matrix
\beq
D_t \eqsp \mathrm{diag} \left\{(z-\bar z) \rho,(1-z) (1-\bar z),z \bar z,(1-z) (1-\bar z),\frac{\delta  (\gamma +2 \rho )}{\delta +\gamma +2 \rho }, z \bar z (\gamma +2 \rho ),(z + \bar z - z \bar z) (\gamma +2 \rho )\right\} \label{rescaleCT}
\eeq
and the core part
\beq
C_t' \eqsp  \left(
\begin{array}{ccccccc}
 2 & 0 & 0 & 0 & 0 & 0 & 0 \\
 0 & -\frac{\rho -\delta }{\delta  (\gamma +\rho ) (\gamma +2 \rho )} & 0 & 0 &
   \frac{1}{\delta +\gamma +2 \rho } & 0 & \frac{1}{\rho } \\
 0 & 0 & \frac{\gamma +\rho }{\delta  (\delta -\rho ) (\gamma +2 \rho )} &
   \frac{1}{\delta  (\gamma +2 \rho )} & \frac{1}{\delta +\gamma +2 \rho } & \frac{\gamma
   +\rho }{\rho  (\delta -\rho )} & 0 \\
 0 & 0 & \frac{1}{\delta  (\gamma +2 \rho )} & \frac{\delta -\rho }{\delta  (\gamma +\rho
   ) (\gamma +2 \rho )} & \frac{1}{\delta +\gamma +2 \rho } & 0 & \frac{1}{\rho } \\
 0 & \frac{1}{\delta +\gamma +2 \rho } & \frac{1}{\delta +\gamma +2 \rho } &
   \frac{1}{\delta +\gamma +2 \rho } & 1 & -\frac{\delta }{\delta +\gamma +2 \rho } &
   \frac{\delta }{\delta +\gamma +2 \rho } \\
 0 & 0 & \frac{\gamma +\rho }{\rho  (\delta -\rho )} & 0 & -\frac{\delta }{\delta +\gamma
   +2 \rho } & \frac{\delta  \rho +2 \gamma ^2+6 \gamma  \rho +3 \rho ^2}{\rho  (\delta
   -\rho )} & -1 \\
 0 & \frac{1}{\rho } & 0 & \frac{1}{\rho } & \frac{\delta }{\delta +\gamma +2 \rho } & -1
   & -\frac{2 \gamma +3 \rho }{\rho } \\
\end{array}
\right) \, .
\eeq
Once again the coordinate dependence factors out!

The  matrices $M_z, \, M_{\bar z}$ for the Pfaffian equations are given in the supplemental material as they are somewhat bulky due to the parameters $\gamma, \, \delta, \, \rho$. We observe that they go into each other under the exchange $z \, \leftrightarrow \, \bar z$ and that they obey the integrability condition
\beq
\partial_z M_{\bar z} - \partial_{\bar z} M_z - M_z M_{\bar z} + M_{\bar z} M_z \eqsp 0 \, . 
\eeq
which guarantees that integrating, say, the $z$ equation first and substituting the result into the $\bar z$ condition will yield a $z$-independent ordinary differential equation on $\bar z$.

Next, let us rescale $\gamma \rar \gamma \, \ep, \, \delta \rar \delta \, \ep, \, \rho \rar \rho \, \ep$ yielding matrices $M_z', \, M_{\bar z}'$. The $\ep$ dependence of these rescaled matrices is linear. What is more, their constant part is diagonal:
\beq
M_z' |_{\ep \rar 0 }\eqsp - \partial_z \, \log \mathrm{diag}\{ z - \bar z, \, 1 - z, \, z, \, 1 - z, \, 1, \,  z,  z + \bar z - z \bar z \}
\eeq
and the same with $z \, \leftrightarrow \, \bar z$. These lowest order equations imply \cite{gluingInter} that our seven masters have the rational denominators
\beq
\underline{d} \eqsp \{ z - \bar z, \, (1-z)(1-\bar z), \, z \bar z, \, (1-z)(1-\bar z), \, 1, \, z \bar z, \, z + \bar z - z \bar z \}
\eeq
exactly as in \eqref{rescaleCT} implying that the second level intersection matrix becomes coordinate independent upon rescaling the master integrals accordingly. On the Pfaffian matrices that induces the transformation
\beq
M_z' \rar N_z \eqsp A^{-1} (M_z'  A - \partial_z  A) \, , \qquad A \eqsp \mathrm{diag}(\underline{d})^{-1} \label{trafo1}
\eeq
and similarly for the $\bar z$ equation. The matrices $N_z, \, N_{\bar z}$ are homogeneous of degree one in $\epsilon$ so that we now have canonical equations w.r.t. this parameter. They can be iterated as in \cite{hennAlg,gluingInter}.

To find the starting values for this process we inspect the singularities of the differential-stripped forms in \eqref{boxBasST}. Asserting that the values of the coordinates $z, \, \bar z$ are sufficiently generic the denominators $p_5, \, p_6$ will have simple poles in the interior of the integration domain (or none). A principal value prescription then assures that these factors do not contribute to the maximal divergence $1/\ep^2$. The situation w.r.t. to $\{s,1-s\}$ and $\{t,1-t\}$ is different: neglecting the factors $p_5, p_6$ the master integrals reduce to the product of two Euler $\beta$ functions depending on $\gamma \, \ep, \, \delta \, \ep$. From here we gain the insight
\beq
\{m_1 \ldots m_7\} \eqsp \left\{0,0,0,0,\frac{1}{\delta^2},0,0 \right\}\frac{1}{\ep^2} + O\left(\ep^{-1}\right) \, . \label{boxLead}
\eeq
The order estimate and the choice of basis \eqref{boxBasST} are again like in the gluing problem \cite{gluingInter}.

The new matrices $N_z, \, N_{\bar z}$ are of dLog type w.r.t. to the arguments $z, \bar z$ respectively and therefore simple to integrate. The set of denominators of $N_z$ is
\beq
\left\{ z, \, z - 1, \, z - \bar z, \, z - \frac{1}{\bar z},  \, z - \frac{\bar z}{\bar z - 1} \right\} \, .
\eeq
Integrating and substituting the result into the $\bar z$ equation, the $z$ dependence must drop at every order in $\ep$ so that we will only meet the denominators $\bar z, \, \bar z - 1$ in that equation. Using Goncharov logarithms \cite{symbolGonchiLog1,symbolGonchiLog2} we obtain a \emph{fibration basis} $G_{\underline{a}}[z] \, G_{\underline{b}}[\bar z]$ in which the weight vector $\underline{a}$ is composed of the letters $\{0,1,\bar z, 1/\bar{z}, \bar{z}/(\bar z - 1) \}$ while $\underline{b}$ consists of the letters $\{0,1\}$. 

However, the double differentiation in formula \eqref{EulerI1} implies that we have to iterate up to $\ep^0$, which means dilogarithm level because the leading order \eqref{boxLead} is a constant over $\ep^2$. Therefore we do not need to introduce generalised polylogarithms; we may simply rely on {\tt Mathematica}'s inbuilt integration routines. At NLO the result is quite concise: we find
\beq
\{n_1 \ldots n_7\}|_{\ep^{-1}} \eqsp \frac{1}{\delta} \left\{ \log\left[\frac{1-z}{1-\bar z}\right], \, 0, \, \log[1- z \bar z], \, 0, \frac{\rho}{\delta} \log[(1-z)(1-\bar z) (1 - z \bar z)], \, 0, \, \log[(1-z)(1-\bar z)] \right\}
\eeq
where $n_i$ denote the pure functions in the seven masters. For the sake of simplicity, integration constants and $i \pi$ from cuts were omitted from the last formula. At dilogarithm level
\begin{eqnarray}
n_1|_{\ep^0} & = &   \frac{2 \, \gamma}{\delta} \left( \text{Li}_2[z] - \text{Li}_2[\bar z)] \right) +\text{Li}_2[z \bar z]-2 \, \text{Li}_2\left[-z \frac{(1-\bar z)}{1-z}\right]+2 \, \text{Li}_2[1-z] \\
&& +\log[1-\bar z] \log[1-z]+\frac{1}{2} \log^2[1-\bar z]-\frac{3}{2} \log^2[1-z]+2 \log[1-z] \log[z] \nonumber
\end{eqnarray}
where we have put $\rho \rar 0$. As a consequence, equation \eqref{EulerI1} yields the desired result \eqref{davyBox1} barring for the cut in $\log[(1-z)/(1-\bar z)]$ since we have only kept track of the leading logarithms in integrating. Note that the symbol of the unphysical $\delta^2$ component of the Euler integral contains the letter $1-z \bar z$, which is absent in the physical part. We will observe the same phenomenon in the double-box computation in the following sections.

\section{The double box by intersection theory} \label{doubleIntersectLong}

In Appendix A we derive the MB representation
\begin{eqnarray}
I_2 & = \frac{1}{2 \, (2 \pi i)^2 \, x_{12}^2} & \int dz_4 \, dz_6 \, u^{z_4} v^{z_6} \, \Gamma[-z_4]^2 \Gamma[-z_6]^2 \Gamma[1+z_4+z_6]^2 * \nonumber \\
 &&  \quad \left( \pi^2 + (\psi[-z_4] - \psi[-z_6])^2 - \psi'[-z_4] - \psi'[-z_6] \right) \label{MBrepI2}
\end{eqnarray}
for the double box. The $\pi^2$ term in the last formula is required to reach numerical agreement with the exact box function \eqref{ExactDB}. In the rest of the article we will construct and solve canonical Pfaffian equations from intersection theory in terms of Goncharov logarithms to match \eqref{ExactDB}, yet for the sake of brevity a comparison is done only on the level of the symbol. We thus drop the $\pi^2$ term from \eqref{MBrepI2} and analyse 
\begin{itemize}
\item the MB representation of the one-loop box with an insertion of $\psi[-z_4] \psi[-z_6]$ into the integrand and
\item the same with an additional $\psi[-z_4]^2 - \psi'[z_4]$ instead, and $u \, \leftrightarrow \, v$ flipped.
\end{itemize}
We can realise the first case writing
\beq
\Gamma[-z_4]^2 \psi[-z_4] \, \Gamma[-z_6]^2 \psi[-z_6] \eqsp \frac{1}{4} \partial_\alpha \partial_\beta \Gamma[-z_4+\alpha]^2 \Gamma[-z_6+\beta]^2 \, , \quad \alpha \eqsp 0 \eqsp \beta
\eeq
Picking residues now goes as in the one-loop example barring for one difference: we close both integration contours to the right. Due to presence of the parameters $\alpha, \beta$ this freezes the MB parameters at $z_4 \eqsp n_4 + \alpha, \, z_6 \eqsp n_6 + \beta, \, n_4,n_6 \in \mathbb{N}_0$. Hence we obtain a modified equation \eqref{sumI1}:
\beq
I_{2,1} \eqsp - \frac{1}{4 \, x_{12}^2} \sum_{n_4,n_6 \eqsp 0}^\infty \partial_\alpha \partial_\beta \partial_\gamma \partial_\delta \, u^{n_4+ \alpha + \gamma} v^{n_6+\beta + \delta} \, \frac{\Gamma[1+n_4+n_6 +\alpha + \beta + \gamma + \delta]^2}{\Gamma[1+n_4+\gamma]^2 \Gamma[1+n_6+\delta]^2} \label{sumI21}
\eeq 
Nevertheless, the modified numerator $\Gamma$ function leads to a substantial complication: both sums are now of $_3F_2$ type so that we obtain a fourfold Euler integral. Index permutations in resolving the two hypergeometric sums into double integrals yield a total of 15 distinct Euler integrals. In order to avoid inflating the size of the basis for intersection theory we prefer formulae with a single coordinate dependent factor, for instance
\beq
I_{2,1} \eqsp - \frac{1}{4 \, x_{12}^2} \partial_\alpha \partial_\beta \partial_\gamma \partial_\delta \, u^{\alpha + \gamma} v^{\beta + \delta} \, J_{2,1} \, ,
\eeq
\beq
J_{2,1} \eqsp \int \frac{dq \, dr \, ds \,dt \, (1 - q)^{-1 + \gamma} \, (1 - r)^\gamma r^{\sigma + \gamma + \delta} \, (1 - s)^{-1 + \delta} \, (1 - t)^{-1 - \sigma} t^{\sigma + \delta} \, \Gamma[1 + \sigma + \gamma + \delta]}{(1 - r - q \, r \, u + q \, r^2 \, u + r \, s \, t \, v)^{1 + \sigma + \gamma + \delta} \, \Gamma[-\sigma] \, \Gamma[-\sigma - \delta] \, \Gamma[\delta] \, \Gamma[1 + \sigma + \delta] \, \Gamma[\gamma])} \, .
\eeq
In the last line $\sigma \eqsp \alpha + \beta$. As in our calculation of the one-loop box, the twist factor $w$ for the intersection theory computation will have to contain two additional terms, here $q^\rho s^\rho$ with a fourth regulator $\rho$. The intersection matrices will depend on $\{\gamma, \, \delta, \, \rho, \, \sigma\}$.

In the second case, we write
\beq
\Gamma[-z_4]^2 (\psi[-z_4]^2 - \psi'[-z_4]) \eqsp - \partial_\alpha ( \partial_\alpha - 2 \partial_\beta) \Gamma[-z_4 + \alpha] \Gamma[-z_4 + \beta] \, , \quad \alpha \eqsp 0 \eqsp \beta \label{differI2}
\eeq
(or the same with $\alpha \, \leftrightarrow \, \beta$). Here we run into an issue about mutiplying distributions: picking residues we would not know whether to assign $z_4 \eqsp n_4 + \alpha$ or $z_4 \eqsp n_4 + \beta$ for integer $n_4$. Likely one can avoid the problem scaling both $\alpha, \, \beta$ by some other parameter and differentiating another time in the latter. To be on the safe side, we rather close the contour to the right for the $z_6$ integration --- so $z_6 \eqsp n_6 \in \mathbb{N}_0$ --- and then for the $z_4$ integration to the left:
\beq
1 + z_4 + n_6 \, = \, - n_4 \, , \quad n_4 \in \mathbb{N}_0 \qquad \Rightarrow \qquad z_4 \eqsp - (1 + n_4 + n_6)
\eeq
Note that {\tt MBresolve.m}'s rule $\{\Re(z_4) \eqsp -0.297642, \, \Re(z_6) \eqsp -0.381832\}$ (cf. Appendix A) puts the real part of $1+z_4+z_6$ to the right of the imaginary axis, so the summation does indeed begin at $n_4 \eqsp 0$. This creates the series
\begin{eqnarray}
I_{2,2} &= & \frac{1}{2 \, x_{12}^2 \, u} \sum_{n_4,n_6 \eqsp 0}^\infty \partial_\alpha (\partial_\alpha - 2 \, \partial_\beta) \partial_\gamma \partial_\delta \, \left(\frac{1}{u}\right)^{n_4+\gamma} \left(\frac{v}{u}\right)^{n_6+\delta} \label{sumI22} \\ && \qquad \qquad \qquad \qquad * \ \ \frac{\Gamma[1+n_4 + n_6 + \alpha + \gamma + \delta] \Gamma[1+n_4+n_6+\beta+\gamma+\delta]}{\Gamma[1+n_4+\gamma]^2 \Gamma[1+n_6+\delta]^2} \, . \nonumber 
\end{eqnarray}
If the Euler integrals do give a re-summation of this series, it will hardly be of importance that its arguments are $1/u, \, v/u$. 
Identifying $\alpha$ and $\beta$ we essentially reproduce the structure of the first case, displaying the latter as a simpler version of the same object. Out of 36 possible Euler representations we pick
\beq
I_{2,2} \eqsp \frac{1}{2 \, x_{12}^2}  \partial_\alpha (\partial_\alpha - 2 \, \partial_\beta) \partial_\gamma \partial_\delta \, u^\beta v^\delta \, J_{2,2} \, ,
\eeq
\beq
J_{2,2} \eqsp \int \frac{dq \, dr \, ds \,dt \, (1 - q)^{-1 + \gamma} \, (1 - r)^{-\alpha + \beta + \gamma} r^{\alpha  + \gamma + \delta} \, (1 - s)^{-1 + \delta} \, (1 - t)^{-1 - \alpha} t^{\alpha+ \delta} \, \Gamma[1 + \beta + \gamma + \delta]}{(- q \, r + q \, r^2 + u - r \, u + r \, s \, t \, v)^{1 + \beta + \gamma + \delta} \, \Gamma[-\alpha] \, \Gamma[-\alpha - \delta] \, \Gamma[\delta] \, \Gamma[1 + \alpha+ \delta] \, \Gamma[\gamma])} \, . \label{J22}
\eeq
Again, to run a reduction by intersection theory an additional term $q^\rho s^\rho$ becomes necessary; in the result for $J_{2,2}$ the limit $\rho \rar 0$ will be well-defined. At intermediate stages, though, we have to cope with all five regularisation parameters.

In the earlier work \cite{gluingInter} the iterative algorithm \cite{inter2} described in Section \ref{introIntersect} has only been tested for the minimal case of two integrations. Will we be able to use it here to deal with a full four Euler parameters? In past tries on direct integration of Euler representations \cite{usFivePoint} it has turned out to be advantageous to integrate out the parameters in the inverse sequence of their appearance in the process of re-summation. Yet, for  the present purpose the sequence $t, \, s, \, r, \, q$ is not too useful, because the coordinate-dependent denominators are quadratic in $r$. Thus we adopt the sequence  $t, \, s, \, q, \, r$ which will entirely linearise the problem.

We begin scrutinising the simpler case $J_{2,1}$. The twist factor $w$ is given by the non-integer part of the powers of the rational factors present in the integral times the extra boundary term $q^\rho s^\rho$. The four-forms can in principle have any numerator and a denominator comprising integer powers of any of the rational factors in the integral. We transform to the coordinates $z, \bar z$ as before, upon which the $r$ integration will linearise in the last step. Calling the coordinate dependent rational factor
\beq
p_9 \eqsp 1 - r  - q \, r \, z \, \bar z+ q \, r^2 \, z \, \bar z + r \, s \, t - r \, s \, t \, z - r \, s \, t \,  \bar z + r \, s \, t \, z \, \bar z
\eeq
a good basis of forms turns out to be
\beq
\left\{\frac{1}{Q \, S \, T \, p_9},\frac{1}{Q \, R \, S \, T},\frac{1}{Q \, R \, s \, T}, \frac{1}{Q \, r \, S \,
   T},\frac{1}{Q \, r \, s \, T},\frac{1}{q \, Q \, p_9},\frac{1}{q \, r \, S \, T},\frac{1}{q \, r \, s \, 
   T},\frac{1}{q \, p_9}\right\} dq \, dr \, ds \, dt \label{basJ21}
\eeq
with $Q \eqsp 1 - q, \, R \eqsp 1-r, \, S \eqsp 1 - s, \, T \eqsp 1 - t$. Assuming $p_9$ to be regular to leading order, the leading divergences of the associated master integrals are
\beq
- \frac{1}{\ep^4 \, \alpha} \left\{0, \, \frac{1}{\gamma^2 \, \delta}, \, \frac{1}{\beta \, \gamma^2}, \, \frac{1}{\gamma \, \delta \,(\alpha +\gamma + \delta)}, \, \frac{1}{\beta \, \gamma \, (\alpha + \gamma + \delta)}, \, 0, \, \frac{1}{\beta \, \delta \, (\alpha + \gamma + \delta)}, \, \frac{1}{\beta^2 \, (\alpha + \gamma + \delta)}, \, 0 \right\} \, .
\eeq
The first master is $J_{2,1}$ itself. The last two masters form a basis for the first step $t$, the last three a basis for the second step $s$, and the last six for the third step $q$. Correspondingly, the connection $\Omega_s$ is a  $2 \times 2$ matrix,  $\Omega_q$ is $3 \times 3$, and finally $\Omega_r$ is $6 \times 6$. We provide the three connections and the four intersection matrices in the ancillary files to this publication.

With nine elements the basis \eqref{basJ21} is astonishingly short. The choice of elements deviates from our intuition about scaling at infinity. Nonetheless, the leading and subleading orders of the expansions around poles are sufficient. What is more, this choice of basis assures homogeneous Euler parameter dependence $s^2 q^2 r^2$ of $C_t$, then $q^2 r^2$ of $C_s$, and finally $r^2$ of $C_q$, while the top level matrix $C_r$ obviously cannot contain any Euler parameter. As before, the coordinate dependence can be scaled out of these intersection matrices taking denominators out of the master integrals. All four $C$ matrices are provided in the ancillary files, too. A second criterion that must be respected in choosing a basis is that the connections $\Omega_{s,q,r}$ do not pick up higher poles and are of pure dLog type, so there should not be a finite remainder once all simple poles are subtracted.

We have not systematically studied the origin of double poles arising from certain candidate masters. In this problem we cannot employ masters with a denominator factor $t$. The effect might be related to the rather special factorisation properties of $p_9$, for instance it factors out $s \, t$ at $r \rar 1$. Yet, a denominator $s$ is not a problem, and $t \rar 0$ or $s \rar 0$ split a factor $R$ off $p_9$ without discriminating against that denominator either; it occurs in master 2.

The Pfaffian matrices have very much the same features as for the one-loop box: rescaling $\gamma \rar \gamma \ep, \, \delta \rar \delta \ep, \, \rho \rar \rho \ep, \, \sigma \rar \sigma \ep$ they are linear in $\ep$ and the leading order equation implies that the masters are pure functions with the denominators
\beq
d_{2,1} \eqsp \{z - \bar z, \, 1, \, 1, \, 1, \, 1, \, (1-z)(1-\bar z), \, 1, \, 1, \, (1-z)(1-\bar z) \} \, .
\eeq
The denominators are taken out by a  transformation as in \eqref{trafo1} on the Pfaffian matrices, upon which the latter are linear in $\epsilon$. Once again, we obtain canonical equations. 

The Pfaffian matrices have the singularities $\{z, \, 1-z, \, \bar z, \, 1 - \bar z, \, z - \bar z, \, 1 - z \bar z, \, z + \bar z - z \bar z\}$ w.r.t. to the coordinates. In this problem, an integration in terms of a fibration basis $G_{\underline{a}}[z] \, G_{\underline{b}}[\bar z]$ with the same alphabet as for the one-loop box --- so $\{0,1,\bar z, 1/\bar{z}, \bar{z}/(\bar z - 1) \}$ in $\underline{a}$ and only $\{0, \, 1 \}$ in $\underline{b}$ --- is not only convenient but even necessary because four logarithm levels are needed due to the fourfold differentiation in formula \eqref{sumI21}. Due to the proliferation of regulator parameters the higher orders of the solution become quite voluminous so that we do not attempt to reproduce them here. Importantly, the limit $\rho \rar 0$ is well-defined for master 1, ie. $J_{2,1}$.

Two comments are in order: first, a Laurent expansion in terms of the regulator parameters is obtained: upon re-instating $\sigma \rar \alpha + \beta$ the logarithm level four part of $I_{2,1}$ (prior to differentiation) contains monomials in $\alpha, \, \beta, \, \gamma, \, \delta$ of total degree four, but also monomials of degree five in $\alpha, \, \beta, \, \delta$ divided by $\gamma$. Second, naively picking the $\alpha \, \beta \, \gamma \, \delta$ component we obtain a fairly concise result with all desired features: it has $u \, \leftrightarrow \, v$ viz $z \, \leftrightarrow \, 1-z, \, \bar z \, \leftrightarrow \, 1-\bar z$ antisymmetry as expected from the denominator $z - \bar z$. The fact that the Laurent expansion does not have a pole in $\delta$ --- which will be a consequence of the fact that we first summed over $n_4$ --- does not spoil this feature. Further, the symbol of the $\alpha \, \beta \, \gamma \, \delta$ component contains only the letters $z, \, 1-z, \, \bar z, \, 1 - \bar z, \, z - \bar z$, whereas the two letters $1-u, \, 1 - v$ do occur in components with $\gamma^n, \, n \, > \, 1$ or $\delta^n, \, n \, > \, 1$. The result for $I_{2,1}$ in terms of Goncharov logarithms and its symbol are included in the ancillary material.

The analysis of the second integral $J_{2,2}$ via intersection theory is strictly analogous, to the extent that we can copy down the basis \eqref{basJ21} with the sole replacement
\beq
p_9 \rar \tilde p_9 \eqsp -q \, r + q \, r^2  + z \, \bar z - r \, z \, \bar z + r \, s \, t - r \, s \, t \, z - r \, s \, t \, \bar z + r \, s \, t \, z \, \bar z \, ,
\eeq
the twist factor $w$ being given by the non-integer part of the powers of the rational factors in \eqref{J22} supplemented by $q^\rho s^\rho$. The computations become more cumbersome because of the fifth regulator parameter; most prominently the inversion of the fourth level intersection matrix $C_r$ takes some effort. 

In this case, too, a Laurent expansion in terms of the regulator parameters arises, and non-surprisingly there is a pole, albeit in $\beta' \eqsp -\alpha + \beta + \gamma$. So how can the right components of this series be selected?

To organise the polylogarithm level 4 part in terms of an independent set of fourth order monomials and fith order monomials over $-\alpha + \beta + \gamma$ we may change to the parameters $\alpha, \, \beta', \, \gamma, \, \delta$ with the ensuing change
\beq
\partial_\alpha \rar \partial_\alpha - \partial_{\beta'} \, , \quad \partial_\beta \rar \partial_{\beta'} \, , \quad \partial_\gamma \rar \partial_\gamma + \partial_{\beta'} \, , \quad \partial_\delta \rar \partial_\delta 
\eeq
which can be used to extract the part of the pure function multiplying $\alpha \beta \gamma \delta$ from the monomials without denominator in the basis with $\beta'$; alternatively we just expand these monomials in terms of the original variables. The action of the derivatives on the fifth order monomials over $\beta'$ would remain to be defined; if we do not Taylor expand in the original variables none of the generated terms is a pure number, suggesting these can simply be discarded.

The problem looks slightly more well-defined using the integral operators
\beq
{\cal O}[x, m] \eqsp \oint \frac{dx}{2 \pi i \, x^{m+1}}
\eeq
to extract the $m$-th power of any parameter. On the fourth order monomials without denominator this is obviously equivalent. Importantly, on all $\alpha^l \gamma^m \delta^n/(\beta - \alpha + \gamma), \, l,m,n \, \geq 0, \, l+m+n \eqsp 5$ present in $J_{2,2}$ we do obtain zero independently of the order of the operations. In fact, out of all 21 possible terms with the composite singularity (by construction, there is no $\beta$ in the numerator) only
\beq
\frac{\delta}{\beta - \alpha + \gamma} \{ \alpha^4, \, \alpha^3 \, \gamma, \, \alpha \, \gamma^3, \, \gamma^4 \} \label{kill1111}
\eeq
are not annihilated by all sequences of the operators $\{{\cal O}[\alpha,1], \, {\cal O}[\beta,1], \, {\cal O}[\gamma,1], \, {\cal O}[\delta,1]\}$, but these structures are absent. In the same vein, the application of $\{{\cal O}[\alpha,2], \, {\cal O}[\gamma,1], \, {\cal O}[\delta,1]\}$ annihilates all 21 structures in any sequence of the operations, while $\{{\cal O}[\beta,2], \, {\cal O}[\gamma,1], \, {\cal O}[\delta,1]\}$ unanimously sends \eqref{kill1111} to $\{-3,-1,0,0\}$ for any sequence of the integral operators and annihilates all others. In conclusion, in the problem at hand our intuition of only looking at the regular monomials is justified. The point will clearly deserve attention in future applications with several singularities like $\beta - \alpha + \gamma$ involving the same parameters.

Denoting the pure function multiplying a monomial as $[\alpha^i \, \beta^j \, \gamma^k \, \delta^l]$, 
\begin{eqnarray}
[\alpha \, \beta \, \gamma \, \delta] & \rightarrow & [\alpha \, \beta' \, \gamma \, \delta] + 2 \, [\alpha \, (\beta')^2 \, \delta] - 2 \,[(\beta')^2 \, \gamma \, \delta] - 6 \, [(\beta')^3 \, \delta] \, , \label{intO1} \\
{[}\alpha^2 \, \gamma \, \delta] & \rightarrow & [\alpha^2 \, \gamma \, \delta] + [\alpha^2 \, \beta' \, \delta] - 2 \, [\alpha \, (\beta')^2 \, \delta] + 3 \, [(\beta')^3 \, \delta] - [\alpha \, \beta' \, \gamma \, \delta ] + [(\beta')^2 \, \gamma \, \delta] \, , \label{intO2} \\
{[}\beta^2 \, \gamma \, \delta] & \rightarrow & [(\beta')^2 \, \gamma \, \delta] + 3 \, [(\beta')^3 \, \delta] \, . \label{intO3}
\end{eqnarray}
Furthermore, ${\cal O}[\alpha,2] \, {\cal O}[\gamma,1] \, {\cal O}[\delta,1]$ and ${\cal O}[\beta,2] \, {\cal O}[\gamma,1] \, {\cal O}[\delta,1]$ do extract the same result from $J_{2,2}$ despite of the different appearance of \eqref{intO2} and \eqref{intO3}. For simplicity, we opt for the latter. Note that the original differential operator $1/2 (\partial_\beta)^2$ in \eqref{differI2} gives 1 on $\beta^2$ and so does ${\cal O}[\beta,2]$. To match the differentiation $(\partial_\beta^2/2 - \partial_\alpha \partial_\beta) \partial_\gamma \, \partial_\delta$ we thus have to take the difference of \eqref{intO3} and \eqref{intO1} which means to build the linear combination
\beq
- [\alpha \, \beta'  \, \gamma \, \delta] - 2 \, [\alpha \, (\beta')^2 \, \delta] + 3 \, [(\beta')^2 \, \gamma \, \delta] + 9 \, [(\beta')^3 \, \delta]  
\label{fishOut}
\eeq
of components of $u^\beta v^\delta J_{2,2}$. Subtracting the same with $u \, \leftrightarrow \, v$ (ie. the $\psi[-z_6]^2 - \psi'[-z_6]$ terms in \eqref{MBrepI2}) and adding in $I_{2,1}$ we reproduce the analytic result for the double box \eqref{ExactDB} up to cuts, as comparing the respective symbols shows.

Last, unphysical components of $J_{2,2}$ --- in particular those with higher positive powers of $\gamma, \, \delta$ --- may contain the letters $1 - u, \, 1 - v$ in their symbol, but those in formulae \eqref{intO1}, \eqref{intO2}, \eqref{intO3} only comprise $\{z, \, 1 - z, \, \bar z, \, 1 - \bar z, \, z - \bar z\}$. We include the function from \eqref{fishOut} and its symbol in the ancillary file as well.

\section{A shortcut to the double box} \label{doubleIntersectShort}

In the preceding section we have learned that it is possible to locate a particular Taylor component of the sum \eqref{sumI21} in the Laurent expansion of the associated Euler integral $J_{21}$. Suppose that this is not a coincidence, but rather a general feature. Now, the sum
\beq
\hat I_{2,1} \eqsp - \frac{1}{4 \, x_{12}^2} \sum_{n_4,n_6 \eqsp 0}^\infty \partial_\alpha \partial_\beta \partial_\gamma \partial_\delta \, u^{n_4+ \alpha + \gamma} v^{n_6+\beta + \delta} \, \frac{\Gamma[1+n_4+n_6 +\alpha + \beta + \gamma + \delta]^2}{\Gamma[1+n_4] \Gamma[1+n_4+2 \, \gamma] \Gamma[1+n_6] \Gamma[1+n_6+2 \, \delta]} \label{hatI21}
\eeq 
has the identical $\alpha \, \beta \, \gamma \, \delta$ component. Yet, the Euler integral from summing over residues is only two-fold because both, the $n_4$ as well as the $n_6$ sums are of $_2F_1$ type. From the two distinct Euler representations found by starting on $\sum n_4$ we choose the one with only one coordinate-dependent factor:
\beq
\hat I_{2,1} \eqsp - \frac{1}{4 \, x_{12}^2} \partial_\alpha \partial_\beta \partial_\gamma \partial_\delta \, u^{\alpha + \gamma} v^{\beta + \delta} \, \hat J_{2,1} \, ,
\eeq
\beq
\hat J_{2,1} \eqsp \int \frac{ds \,dt \, (1 - s)^{2 \, \gamma} s^{\sigma + \gamma +\delta} \, (1 - t)^{-1 - \sigma + \gamma + \delta} t^{\sigma - \gamma + \delta} \, \Gamma[1 + \sigma + \gamma + \delta]}{(1 - s - s \, u + s^2 \, u + s \, t \, v)^{1 + \sigma + \gamma + \delta} \, \Gamma[1 + \sigma -\gamma + \delta] \, \Gamma[-\sigma + \gamma - \delta] \, \Gamma[-\sigma + \gamma + \delta]}
\eeq
with $\sigma \eqsp \alpha + \beta$. There is no need for any additional regulator because all necessary factors are present and carry non-integer parts in their exponents. The intersection theory problem for the new integral is much simpler than before: there are only two variables to integrate out, and the matrices involve only three regulator parameters. Defining
\beq
p_5 \eqsp 1 - s - s \, z \, \bar z + s^2 \, z \, \bar z + s \, t - s \, t \, z - s \, t \, \bar z  + s \, t \, z \, \bar z
\eeq
a suitable basis is
\beq
\left\{ \frac{1}{T \, p_5}, \, \frac{1}{S \, T}, \, \frac{1}{s \, T}, \, \frac{1}{p_5} \right\} ds \, dt \label{basHatI21}
\eeq
where the last two elements are a basis for the inner integration, for which we select the variable $t$. The problem linearises at the second level, like it did for the $r$ integration that we tackled last in the previous section. We list the intersection matrices, the second level connection, and the Pfaffian matrices in Appendix B. Conveniently, there are only the five letters $\{z, \, 1 - z, \, \bar z, \, 1 - \bar z, \, z - \bar z \}$ so that a fibration basis $G_{\underline{a}}[z] G_{\underline{b}}[\bar z]$ is sufficient in which the weight vector $\underline{a}$ has entries from the smaller set $\{0,\, 1, \, \bar z\}$, and $\underline{b}$ is composed of $\{0, \, 1\}$ again.

Rescaling $\sigma \rar \sigma \ep, \, \gamma \rar \gamma \ep, \, \delta \rar \delta \ep$ the equations are linear in $\ep$. Once again, the constant order implies that the master integrals have rational denominators:
\beq
\hat d_{2,1} \eqsp \{z-\bar z, \, 1, \, 1, \, (1-z)(1-\bar z) \}
\eeq
Transforming to pure functions we obtain canonical equations w.r.t. to $\ep$ whose iteration starts on
\beq
\frac{1}{(-\sigma + \gamma + \delta) \, \ep^2} \, \left\{0, \, \frac{1}{2 \, \gamma}, \, \frac{1}{\sigma + \gamma + \delta}, \, 0 \right\} \, .
\eeq
We remark that the order counting is similar to the more complicated case in the last section: the integral $\hat J_{2,1}$ has two denominator $\Gamma$ functions singular in $\epsilon$ (not four). Scaling up the result for master 1 by the factors $\ep^2 (\sigma - \gamma + \delta)(\sigma - \gamma - \delta)$, the polylogarithm level 4 once again comes with monomials of total order four in the regulator parameters, or of total order five, but divided by $\gamma$. For the $\alpha \, \beta \,  \gamma \, \delta$ component of $\hat I_{2,1}$ we find exactly the same symbol as for $I_{2,1}$, while the entire calculation requires nearly no computing resources!

We can modify $I_{2,2}$ in a similar fashion:
\begin{eqnarray}
\hat I_{2,2} &= & \frac{1}{2 \, x_{12}^2 \, u} \sum_{n_4,n_6 \eqsp 0}^\infty \partial_\alpha (\partial_\alpha - 2 \, \partial_\beta) \partial_\gamma \partial_\delta \, \left(\frac{1}{u}\right)^{n_4+\gamma} \left(\frac{v}{u}\right)^{n_6+\delta} \label{hatSumI22} \\ && \qquad \qquad \quad \ * \ \ \frac{\Gamma[1+n_4 + n_6 + \alpha + \gamma + \delta] \Gamma[1+n_4+n_6+\beta+\gamma+\delta]}{\Gamma[1+n_4] \Gamma[1+n_4+2 \, \gamma] \Gamma[1+n_6] \Gamma[1+n_6+2 \, \delta]} \, . \nonumber 
\end{eqnarray}
Summing over $n_4$ first we find four Euler representations, of which we pick
\beq
\hat I_{2,2} \eqsp \frac{1}{2 \, x_{12}^2}  \partial_\alpha (\partial_\alpha - 2 \, \partial_\beta) \partial_\gamma \partial_\delta \, u^\beta v^\delta \, \hat J_{2,2} \, ,
\eeq
\beq
\hat J_{2,2} \eqsp \int \frac{ds \,dt \, (1 - s)^{-\alpha + \beta + 2 \, \gamma} s^{\alpha + \gamma + \delta} \, (1 - t)^{-1 - \alpha + \gamma + \delta} t^{\alpha - \gamma + \delta} \, \Gamma[1 + \beta + \gamma + \delta]}{(-s + s^2 + u - s \, u + s \, t \, v)^{1 + \beta + \gamma + \delta} \, \Gamma[1 + \alpha - \gamma + \delta] \Gamma[-\alpha + \gamma - \delta] \Gamma[-\alpha + \gamma + \delta]} \, .
\eeq
Again, this is a two-parameter case with one regulator parameter less than before, so very much easier to handle than $I_{2,2}$. We can use the basis \eqref{basHatI21} replacing
\beq
p_5 \rar \tilde p_5 \eqsp -s  + s^2 + z \, \bar z - s \, z \, \bar z  + s \, t - s \, t \, z - s \, t \, \bar z + s \, t \, z \, \bar z \, .
\eeq
The same denominators are found as for \eqref{basHatI21} itself. Transforming to pure functions results in canonical equations. Their iteration starts with the second order poles
\beq
\frac{1}{(-\alpha + \gamma + \delta) \, \ep^2} \, \left\{0, \, \frac{1}{-\alpha + \beta + 2 \, \gamma}, \, \frac{1}{\alpha + \gamma + \delta}, \, 0 \right\} \, .
\eeq
The remaining considerations about $I_{2,2}$ go through verbatim; note, though, that the complicated singularity is a simple pole in $\beta - \alpha + 2 \, \gamma$ in this case. The factor 2 on $\gamma$ causes a rescaling of some of the coefficients in \eqref{intO1},\eqref{intO2},\eqref{intO3} so that \eqref{fishOut} becomes
\beq
- [\alpha \, \beta'  \, \gamma \, \delta] - 4 \, [\alpha \, (\beta')^2 \, \delta] + 3 \, [(\beta')^2 \, \gamma \, \delta] + 18 \, [(\beta')^3 \, \delta]
\eeq
Nonetheless, subtracting the same at $z \rar 1-z, \, \bar z \rar 1 - \bar z$ we reproduce the total $I_{2,2}$ contribution. 

To illustrate the comparative simplicity of this calculation we display the $N_z$ matrix for the more complicated integral $\hat I_{2,2}$ (i.e. the re-definition of the regulator parameters by $\ep$ has been done, and the constant order has been removed taking the respective denominators out of the master integrals). It is convenient to decompose the Pfaffian matrix $N_z$ into three parts pertaining to the coordinate-dependent denominators $z, \, z-1, z - \bar z$:
\beq
N_z \eqsp \frac{1}{z} \, N_{z,0} + \frac{1}{z-1} N_{z,1} + \frac{1}{z-\bar z} N_{z,\bar z} \, ,
\eeq
where
\beq
N_{z,0} \eqsp \left(
\begin{array}{cccc}
 \alpha  & -\frac{\alpha  \beta + \gamma^2-\delta ^2}{\beta +\gamma + \delta } &
   -\frac{\beta  (\alpha + \gamma + \delta)}{\beta +\gamma+\delta} & -\beta  \\[1 mm]
 0 & 0 & 0 & 0 \\[1 mm]
 \frac{1}{2} (\beta + \gamma + \delta) & \frac{1}{2} (\alpha -\beta -2 \gamma ) &
   \frac{1}{2} (\alpha -2 \beta -\gamma -\delta ) & -\frac{1}{2} (\beta +\gamma+\delta)
   \\[1 mm]
 \frac{1}{2} (\alpha -\gamma-\delta) & -\frac{(\alpha +\beta +2 \delta ) (\alpha - \gamma
   -\delta )}{2 (\beta +\gamma+\delta)} & -\frac{(\alpha -\gamma-\delta )
   (\alpha +\gamma+\delta)}{2 (\beta +\gamma+\delta)} & \frac{1}{2} (\alpha -2 \beta
   +\gamma+\delta)
\end{array}
\right) \, ,
\eeq
\beq
N_{z,1} \eqsp \left(
\begin{array}{cccc}
 \frac{1}{2} (\alpha +\beta -2 \delta ) & \frac{(\alpha +\beta -2 \delta ) (\alpha +\beta
   +2 \delta )}{2 (\beta +\gamma+\delta)} & \frac{(\alpha +\beta -2 \delta ) (\alpha
   +\gamma + \delta )}{2 (\beta +\gamma+\delta)} & -\frac{1}{2} (\alpha -\beta +2 \gamma
   ) \\[1 mm]
 -\frac{1}{2} (\beta +\gamma+\delta) & -\frac{1}{2} (\alpha +\beta +2 \delta ) &
  - \frac{1}{2} (\alpha +\gamma+\delta) & -\frac{1}{2} (\beta +\gamma +\delta  ) \\[1 mm]
 0 & 0 & 0 & \beta + \gamma + \delta  \\[1 mm]
 0 & 0 & 0 & -2 \delta 
\end{array}
\right) \, ,
\eeq
\beq
N_{z,\bar z} \eqsp \left(
\begin{array}{cccc}
 -2 (\alpha +\beta ) & 0 & 0 & 0 \\
 0 & 0 & 0 & 0 \\
 0 & 0 & 0 & 0 \\
 0 & 0 & 0 & 0 \\
\end{array}
\right) \, . \label{explainSigns}
\eeq
Up to signs, the matrix $N_{\bar z}$ is obtained from $N_z$ simply mapping $z \, \leftrightarrow \, \bar z$. As the denominator of the first master is odd under this replacement and the other three are even, we must flip the sign of the $\{1,i\}$ and $\{i,1\}$ entries for $i \, > \, 1$.

\section{Conclusions}

The conformal off-shell ladder diagrams in four dimensions have first been computed in \cite{davydychev1,davydychev2}. In this article we have recovered the known expressions for the one- and two-loop off-shell ladder diagrams (viz the box and the double box) in exactly four dimensions in a different way that is a priori more involved but should hopefully be of wider applicability: if a sufficiently simple \emph{Mellin-Barnes representation} of a Feynman graph can be constructed, one may close its integration contours on the left or right half-plane and pick residues, thereby translating the integral into a multiple sum (or several such sums). At least in our two examples --- but likely for all Feynman graphs --- the sum over any individual counter is of hypergeometric $_{p+1}F_p$ type. Replacing the function by its $p$-fold \emph{Euler integral representation} and assuming that the integrations can be swapped over the ensuing sums we can iterate the process because the Euler integrand consists of the same elements as the original summand, so rational functions to some power and $\Gamma$ functions \cite{usFivePoint}. The number of integrations depends on the sequence of the counters being summed over: with a bit of luck many of the steps yield low-dimensional integrals or even geometric sums. The final multiple Euler integrals can at times directly be evaluated in terms of multiple polylogarithms.

It is possible to incorporate polygamma functions into the formalism --- by way of example  $\Gamma[z] \, \psi[z] \eqsp \partial_\alpha|_{\alpha \eqsp 0} \, \Gamma[z+\alpha]$ --- and leaving that differentiation to the very end assuming that it commute over both, summations and integrations. Similarly, higher poles in the Mellin-Barnes representation may be dealt with introducing further parameters. The re-summation into Euler integrals is apparently not hindered by such \emph{regulator parameters} as we call them in the main text, although their presence may significantly increase the number of integrations. 

In general, expanding the Euler integrands in these parameters and integrating afterwards is not a well-defined procedure, as our two examples show. Therefore one is faced with the problem of solving rather general Euler integrals, see eg. \eqref{EulerI1} already for the off-shell one-loop box. To this end we employ \emph{intersection theory} for hypergeometric functions. This allows us to derive \emph{Pfaffian systems} of first order differential equations on the Euler integrals. The latter are apparently always of \emph{canonical form} \cite{hennAlg} implying that one can solve them in terms of \emph{iterated integrals}.

From a two-parameter MB representation of the one-loop box \cite{davydychev1} we found a two-parameter Euler representation. We studied the latter by bivariate intersection theory \cite{inter2}. For the double box we were able to construct a two-parameter MB representation \eqref{MBrepI2} as well, based on the Cheng-Wu trick \cite{ChengWu} and heavy use of the {\tt barnesRoutines.m} package \cite{kosower} to integrate out superfluous MB parameters by Barnes' lemmata. Several four-fold parametric derivatives of sums over residues now need to be determined. We first took these sums as they arise from picking residues and translated them into four-fold Euler integrals. Following the algorithm of \cite{inter2} the associated four-layer intersection theory problem can indeed be solved and integrated, and we confirm the known answer for the double box \cite{davydychev1}. On the one hand this is quite spectacular to see, on the other the computational effort is considerable. The most complicated step is the inversion of the matrix of level four intersection numbers, a dense $9 \times 9$ matrix containing five regulator parameters. 

Yet, we are not really interested in the full, undifferentiated sum of residues including the regulators, but rather in specific Taylor components. It is possible to shift the two parameters affecting the denominator $\Gamma$ functions in \eqref{sumI21}, \eqref{sumI22} without changing the Taylor components in question. This yields a bivariate intersection theory problem with one regulators less which can be solved effortlessly yielding the same result. In this example --- just as intended in \cite{inter2} --- intersection theory very efficiently replaces the IBP technique in decomposing integrals in terms of a basis of masters, and the ensuing construction of Pfaffian differential equations.

Prospectively, many steps of these computations can be automated, e.g. the construction of series of residues for multi-dimensional MB representations, cf. \cite{conicHulls}, and selecting the best suited of the associated Euler integrals. This paper thus possibly hints at a new approach to the evaluation of Feynman graphs. A pivotal input is a fairly low-dimensional MB representation, which is, of course, hard to obtain.

A problem more inherent to intersection theory is the construction of bases. For $n$-parameter Euler integrals the basic  idea is to choose monomials $1/(p_1 ... p_n)$ involving any $n$ rational factors present in the Euler integrand. The monomials can have more or less constituents if some of their factors involve several integration variables; we may also add simple numerators in that case. As yet there is no definite prescription. Nevertheless, with a little hindsight one quickly finds a basis so that
\begin{itemize}
\item only the lowest layers of the Taylor expansion around any pole are solicited,
\item the connections beyond the first level do not contain higher poles,
\item a rescaling of the masters by denominators depending on the kinematic invariants/outer variables makes the intersection numbers constant, i.e. their core part only depends on the regulator parameters. As a test, their determinants should cleanly fall apart into factors involving outer variables and others consisting of regulators.
\end{itemize}
Rescaling all regulator parameters by a common scale $\ep$ (which acts as a dimensional regulator) a good choice of basis implies that the equations will be canonical w.r.t. $\ep$ and can be solved in terms of iterated integrals\footnote{In preliminary studies we have occasionally seen weight drop on some masters, which is related to off-diagonal elements at $\ep^0$ in the Pfaffian matrices.}. In our two examples Goncharov polylogarithms \cite{symbolGonchiLog1,symbolGonchiLog2} are sufficient. The result is a Laurent series in terms of the regulators. Picking the desired components we match the expressions in \cite{davydychev1,davydychev2}. 

However, singularities in a linear combination of the regulators complicate this step. While there is a definite answer for the double box it will be interesting to see how this problem can be handled in more complicated situations. Our scheme is definitely operational, yet it would be good to establish why the Euler integrals we study can be singular in these parameters while the series from which they originate are apparently regular.

Finally, the MB representation of any Feynman graph can be translated into an A-hypergeometric system \cite{beukers, chinese4}. What is more, it is possible to construct a Pfaffian system starting from the ideal of differential operators annihilating the underlying generalised hypergeometric function \cite{chestnov1,chestnov2}. In general, the resulting partial differential equations are not of canonical form, though. Both approaches start from the MB representation --- can we pinpoint a link between the methods? While mathematically more advanced techniques are put forward e.g. in \cite{chestnov3} we emphasize that our computations are not restricted to any cut; they rather give a full result.

\section{Acknowledgements}

The author is a fellow of the Heisenberg scheme of Deutsche Forschungsgemeinschaft (DFG), grant agreement ED 78/7-1 or 441791296. We are grateful to J.~Bl\"umlein for help on some references, and to V.~Smirnov for comments on the manuscript. A.~Davydychev provided feedback on the MB representations used in \cite{davydychev1,davydychev2}.

\section*{Appendix A: MB representations for the one- and two-loop ladders}

As the two box integrals are finite we work in exactly four dimensions. Further, since they are both conformal we may amputate the propagators attached to one outer point without loss of information. To this end we send the bottom point to infinity. In case of the one-loop box integral this leaves the propagators 
\beq
\frac{1}{x_{15}^2 x_{25}^2 x_{35}^2} \eqsp 2 \int \frac{dq_1 dq_2 dq_3 \, \delta(1 - q_1 - q_2 - q_3)}{(q_1 \, x_{15}^2 + q_2 \, x_{15}^2 + q_3 \, x_{35}^2)^3}
\eeq
employing the standard Feynman-parameter trick. Completing squares and integrating out the ``loop-momentum" $x_{15} - q_2 x_{12} - q_3 x_{13}$ we find
\beq
I_1 \eqsp \int_0^1 \frac{dq_1 dq_2 dq_3 \, \delta(1 - q_1 - q_2 - q_3)}{q_1 q_2 \, x_{12}^2 + q_1 q_3 \, x_{13}^2 + q_2 q_3 \, x_{23}^2} \, .
\eeq
To derive a Mellin-Barnes (MB) representation, see \cite{theMB} and references therein, we use the defining relation
\beq
\frac{1}{(A+B)^\lambda} \eqsp \frac{1}{2 \pi i \Gamma[\lambda]} \int_{-i \infty}^{i \infty} dz \frac{B^z}{A^{z+\lambda}} \Gamma[-z] \Gamma[z +\lambda] \label{defMB}
\eeq
twice to arrive at
\begin{eqnarray}
I_1 &=& \frac{1}{(2 \pi i)^2 \, x_{12}^2} \int dz_1 dz_2 \, u^{z_1} v^{z_2} \, \Gamma[-z_1] \Gamma[-z_2] \Gamma[1+z_1+z_2] \, * \\ && \quad \int_0^1 dq_1 dq_2 dq_3 \, \delta(1 - q_1 - q_2 - q_3) \, q_1^{-z_1-1} q_2^{-z_2-1} q_3^{(1+z_1+z_2)-1} \, ,
\qquad u \eqsp \frac{x_{13}^2}{x_{12}^2} \, , v \eqsp \frac{x_{23}^2}{x_{12}^2} \, . \nonumber
\end{eqnarray}
Finally, the Feynman parameters can be integrated out thanks to \cite{ambre}
\beq
\int_0^1 \prod dq_i \, q_i^{n_i-1} \, \delta(1 - \sum_i q_i) \eqsp \frac{\prod_j \Gamma[n_j]}{\Gamma\bigl[\sum_k n_k\bigr]} \label{ambreInt}
\eeq
yielding the pretty expression \cite{davydychev1}
\beq
I_1 \eqsp \frac{1}{(2 \pi i)^2 \, x_{12}^2} \int dz_1 dz_2 \, u^{z_1} v^{z_2} \, \Gamma[-z_1]^2 \Gamma[-z_2]^2 \Gamma[1+z_1+z_2] ^2 \, , \notag
\eeq
cf. equation \eqref{MBrepI1} in the main text.

\newpage

At this point we have transformed the original loop integral over one four-momentum into a parametric double integral. Did we make progress, though? One-dimensional MB integrals can often be computed exactly, but already the simple looking expression for $I_1$ resists naive attempts on integration. Yet, as the $\Gamma$ function exponentially decays in the imaginary direction the method does allow for a fairly stable numerical evaluation of Feynman integrals \cite{MBm}. However, if straight-line contours are desired, their positions along the real axis must be so arranged that they run between $\Gamma$ functions with positive and negative sign, respectively, of the MB parameters $z_i$. This may necessitate jumping over a singularity and thereby extracting some residues. The result is a sum over --- partially lower-dimensional --- MB integrals. For planar Feynman integrals the two most commonly applied strategies for fixing reals parts are realised in the packages {\tt MB.m} \cite{MBm} and {\tt MBresolve.m} \cite{MBresm}, where the latter can be uploaded into the former.

Next, we turn to the double box. Mellin-Barnes representations can be built loop-wise: we send the bottom point to infinity also for the double box, thereby amputating the two propagators attached to it. We could substitute the MB representation \eqref{MBrepI1} for the remaining three propagators attached to the left vertex in the right panel of Figure 1, and take the product with the two propagators on the top right and on the right of the graph. The process described in the opening paragraphs of this Appendix is then repeated. The six point permutations of \eqref{MBrepI1} generate six slightly inequivalent-looking four-parameter representations for the full two-loop integral, cf. \cite{davydychev1}. Picking residues of four simultaneous contour integrals can be cumbersome --- there will usually be several series of residues, cf. the {\tt  MBconicHulls.m} package \cite{conicHulls} --- whereas the otherwise very efficient {\tt barnesRoutines.m} \cite{kosower} does not spot ways to apply Barnes' lemmata to integrate any of the four MB parameters out of these six representations.

An alternative is to put together all five propagators in a single step employing five Feynman parameters, and generating one common denominator
\beq
q_1 x_{15}^2 + q_2 x_{26}^2 + q_3 x_{56}^2 + q_4 x_{35}^2 + q_5 x_{36}^2 \, .
\eeq
Successively integrating over both (dual) loop-momenta (ie. the positions of the two vertices) the integral representation
\beq
I_2 \eqsp \int \frac{dq_1 \ldots dq_5 \, \delta(1 - \sum_i q_i)}{U(q_1 \ldots q_5) \, f(q_1 \ldots q_5, x_{12}^2, x_{13}^2, x_{23}^2)}
\eeq
is obtained, where $U$ and $f$ are the famous Symanzik polynomials. At first glance this looks hardly helpful: $U$ has six terms and $f$ even nine, so to split both into monomials we apparently need the astounding number of 13 MB parameters. Nonetheless, appealing to a trick \cite{Nakanishi,ChengWu,Bluemlein} we will only need seven. What is more, {\tt barnesRoutines.m} will be able to integrate out five of these, leading to a two-parameter representation looking like a higher-transcendentality analogue of \eqref{MBrepI1}.

To this end we may split the parameters into the groups $\{q_1,q_3,q_4\}$ and $\{q_2,q_5\}$, respectively. This partition has a graphical origin: the propagators $1/x_{26}^2, \, 1/x_{36}^2$ associated with $q_2$ and $q_5$, respectively, are not connected to the left vertex in the right panel of Figure 1. This can be used to simplify the Symanzik polynomials, the structure of which is also closely associated with the Feynman graph. Now, in one of the partitions the integration range for the parameters is extended from $[0,1]$ to $[0,\infty]$, upon which they can be scaled out of the $\delta$ function \cite{ChengWu,Bluemlein}. We choose $q_2, q_5$ as these \emph{Cheng-Wu variables} upon which we can put
\beq
q_1 + q_3 + q_4 \rar 1 
\eeq
in the polynomials. Then
\beq
U \eqsp q_3 (q_1+q_4) + q_2 + q_5
\eeq
and
\begin{eqnarray}
f \eqsp q_1 q_2 q_3 x_{12}^2 + q_1 (q_2 q_4 + q_3 q_4 + q_3 q_5 + q_4 q_5) x_{13}^2 + q_2 (q_3 q_4 + q_5) x_{23}^2 \, .
\end{eqnarray}
The idea is now to decompose $f$ into monomials introducing six MB parameters, which we will do from the back to the front. The $U$ polynomial is treated differently: we can use the formula
\beq
\int_0^\infty dq \, q^m (q + A)^n \eqsp A^{1+m+n} \frac{\Gamma[1+m] \Gamma[-1-m-n]}{\Gamma[-n]}
\eeq
to eliminate the Cheng-Wu variables $\{q_2, q_5\}$ in favour of $\Gamma$ functions. Having done so, the remainder of the $U$ polynomial unfortunately still counts two terms, namely $q_3 (q_1+q_4)$, which we split introducing a seventh MB parameter. In a final step, equation \eqref{ambreInt} can be used to integrate out $\{q_1, q_3, q_4\}$ resulting into a seven-parameter MB representation.

We submit this to {\tt MBresolve.m} to fix the position of the straight line contours parallel to the imaginary axis, and then to {\tt barnesRoutines.m}\footnote{The package incorporates corollaries of Barnes' lemmata developed for the evaluation of MB integrals in \cite{BDS}.}. Amazingly, it achieves a reduction to a list of two- and three-parameter integrals. These come in three groups:
\begin{itemize}
\item six two-parameter integrals with real parts fixed at $\{\Re(z_4) \eqsp -0.297642, \, \Re(z_6) \eqsp -0.381832\}$.
\item Two three-parameter integrals at $\{\Re(z_4) \eqsp -0.297642, \, \Re(z_5) \eqsp 0.0186742, \, \Re(z_6) \eqsp -0.381832\}$. The sum of their integrands --- which contains a difference of digamma functions --- can be written as
\begin{eqnarray}
&& - \, \partial_\delta \, \Gamma[-\delta - z_4 - z_5]  \Gamma[\delta + z_5 - z_6] * \\
&& \ \ \, \frac{u^{z_4} v^{z_6} \Gamma[-z_4]^2 \Gamma[1 + z_4 + z_5]\Gamma[-z_6]^2 \Gamma[1 + z_4 + z_6] \Gamma[
  1 - z_5 + z_6] \Gamma[-z_4 - z_5 + z_6]}{x_{12}^2  \Gamma[-z_4 - z_6] \Gamma[1 - z_4 - z_5 + z_6]} \nonumber
\end{eqnarray}
at $\delta \eqsp 0 $. Since $\delta$ is supposed to be infinitesimal we additionally supply the rule $\delta \rar 1/100$ to make the package work and run {\tt barnesRoutines.m} another time on this integral. It eliminates $z_5$ by Barnes' second lemma yielding a sum of three parts. Adding the $O(\delta^1)$ terms of the Tayler expansion of their sum to the six terms mentioned in the first bullet point and simplifying by \eqref{doubleGamma} and it's corollaries --- notably $\psi[-z] - \psi[1+z] \eqsp \pi \cot[\pi z]$ and its derivative --- we obtain a very concise MB representation of the off-shell double box:
\begin{eqnarray}
I_2 & = \frac{1}{2 \, (2 \pi i)^2 \, x_{12}^2} & \int dz_4 \, dz_6 \, u^{z_4} v^{z_6} \, \Gamma[-z_4]^2 \Gamma[-z_6]^2 \Gamma[1+z_4+z_6]^2 * \nonumber \\
 &&  \quad \left( \pi^2 + (\psi[-z_4] - \psi[-z_6])^2 - \psi'[-z_4] - \psi'[-z_6] \right) \nonumber
\end{eqnarray}
which is equation \eqref{MBrepI2} in the main text. It is not explicitly stated in \cite{davydychev1,davydychev2} but it should be closely related to the formulae in those works. Its form is quite suggestive --- do higher off-shell ladder integrals also possess a two-parameter  MB  representation in which polygamma functions are inserted into that of the one-loop box? 
\item We have tacitly omitted another group of four terms for the reason that they add up to zero:
\begin{eqnarray}
&& \int dz_4 \, dz_6 \, u^{z_4} v^{z_6} \, \frac{\Gamma[-z_4]^2 \Gamma[-z_6] \Gamma[1 + z_4 + z_6] \Gamma[1 + z_6 - z_7] \Gamma[-z_7] \Gamma[1 + z_4 + z_7] \Gamma[-z_6 + z_7]}{z_4-z_6+z_7} * \nonumber \\
&& \qquad \frac{\psi[-z_4] - \psi[-z_6 + z_7]}{(2 \pi i)^2 \, x_{12}^2} \label{annoy}
\end{eqnarray}
evaluated at $\{\Re(z_4) \eqsp -0.617028, \, \Re(z_6) \eqsp -0.379074, \, \Re(z_7) \eqsp -0.148157\}$ minus the same at
$\{\Re(z_4) \eqsp -0.297642, \, \Re(z_6) \eqsp -0.381832, \, \Re(z_7) \eqsp -0.0672329\}$. None of the real parts of the arguments of the $\Gamma$ or $\psi$ functions in \eqref{annoy} changes sign upon switching from one prescription to the other, but the denominator $z_4-z_6+z_7$ does. Hence the difference of the two evaluations is given by the residue at $z_4-z_6+z_7 \eqsp 0$. However, at that point the difference of $\psi$ functions in \eqref{annoy} vanishes.
\end{itemize}
{\tt Mathematica}'s inbuilt {\tt NIntegrate[]} routine is sufficient to numerically verify these statements: in single precision the three-parameter integrals do not take too much time and seem to deviate from a putative exact result in the fifth digit as we may observe on the cancellation of the four terms in the third bullet point. It is indicative that the failure to strike the exact outcome zero is O(1) and hence as large as an imaginary part accumulated by the numerical integration.

The differentiation trick in the second bullet point can be checked to very good precision, too. Last, in single precision the two-parameter integrals in  \eqref{MBrepI2} perfectly agree\footnote{The ratio comes out as 1. + 0. i.} with the analytic result \eqref{ExactDB} for the points $\{u,v\} \eqsp \{1/2,1/3\}, \, \{1/2,3/2\}, \, \{3/2,5/4\}, \, \{3/2,1/5\}$. However, even putting the option {\tt WorkingPrecision $\rar$ 10} the numerical integrations fail to come to an end\footnote{In this attempt we rationalised the rules for the real parts.}.

\section*{Appendix B}

We start listing quantities relevant to the evaluation of $\hat I_{2,1}$. The first and second level intersection matrices are
\beq
C_t \eqsp - \frac{1}{(\sigma + \gamma - \delta) \, s^2} \, D_t^{-1} 
\left(\begin{array}{cc}
 \frac{2 \gamma }{\sigma-\gamma-\delta } & 1 \\
 1 & -\frac{2 \delta }{\sigma +\gamma + \delta} \\
\end{array}
\right) \, D_t^{-1} \, , \qquad D_t \eqsp \mathrm{diag}\{1, \, (1-z)(1-\bar z)\} \, ,
\eeq
\beq
C_s \eqsp - \frac{1}{(\sigma + \gamma - \delta) (\sigma - \gamma - \delta)} \, D_s^{-1}
\left(
\begin{array}{cccc}
 \frac{2 (\sigma+\gamma-\delta)}{\sigma+\gamma+\delta} & 0 & 0 & 0 \\
 0 & 0 & -1 & 1 \\
 0 & -1 & \frac{2 \left(\sigma^2-\gamma^2-\delta^2 \right)}{(\sigma+\gamma-\delta)
   (\sigma+\gamma+\delta)} & -\frac{2 (\sigma -\delta )}{\sigma+\gamma-\delta} \\
 0 & 1 & -\frac{2 (\sigma -\delta)}{\sigma+\gamma-\delta} & \frac{4 \delta  (\sigma-\delta)}{(\sigma+\gamma-\delta ) (\sigma+\gamma+\delta)}
\end{array}
\right) \, D_s^{-1} \, , 
\eeq
with
\beq
D_s \eqsp \mathrm{diag} \{z - \bar z, \, 1, \, 1 , (1-z)(1-\bar z) \} \, .
\eeq
It is simplest to write out the second level connection residue by residue:
\beq
\Omega_s \eqsp \frac{1}{s} \, \Omega_{s,0} + \frac{1}{s-1} \, \Omega_{s,1} + \frac{1}{s-1/z} \, \Omega_{s,1/z} + \frac{1}{s-1/\bar z} \, \Omega_{s,1/\bar z} + \frac{1}{s-1/(z \bar z)} \, \Omega_{s,1/(z \bar z)}
\eeq
where
\beq
\Omega_{s,0} \left(
\begin{array}{cc}
\sigma+\gamma+ \delta -1 & (1-z) (1-\bar z) (\sigma + \gamma + \delta) \\
 0 & \sigma+\gamma-\delta -1 \\
\end{array}
\right) \, , \qquad \Omega_{s,1} \eqsp \left(
\begin{array}{cc}
 2 \gamma  & 0 \\
 \frac{\sigma-\gamma-\delta}{(1-z) (1-\bar z)} & 0 \\
\end{array}
\right) \, , 
\eeq
\beq
\Omega_{s,1/z} \eqsp \Omega_{s,1/\bar z} \eqsp - \left(
\begin{array}{cc}
 \sigma + \gamma + \delta & (1-z)(1-\bar z) (\sigma + \gamma + \delta) \\
 \frac{\sigma -\gamma-\delta }{(1-z)(1-\bar z)} & \sigma - \gamma -\delta \\
\end{array}
\right) \, , \quad
\Omega_{s,1/(z \bar z)} \eqsp \left(
\begin{array}{cc}
 0 & 0 \\
 \frac{\sigma-\gamma-\delta}{(1-z)(1-\bar z)} & -2 \gamma  \\
\end{array}
\right) \, .
\eeq
Scaling up by the same denominators $ \{z - \bar z, \, 1, \, 1 , (1-z)(1-\bar z) \}$ as in $D_s$ the Pfaffian matrices transform according to \eqref{trafo1}. Like in Section \ref{introIntersect} the notation $N_z, \, N_{\bar z}$ refers to this stage of the chain of transformations:
\beq
N_z \eqsp \frac{1}{z} \, N_{z,0} + \frac{1}{z-1} N_{z,1} + \frac{1}{z-\bar z} N_{z,\bar z} \, ,
\eeq
\beq
N_{z,0} \eqsp \left(
\begin{array}{cccc}
 \sigma -\gamma  & \frac{\sigma^2 - \gamma^2 - \delta ^2}{\sigma + \gamma + \delta} & -\delta  & -\sigma -\gamma \\[1 mm]
 -\frac{1}{2} (\sigma + \gamma + \delta) & -\gamma  & \phantom{-} \frac{1}{2} (\sigma + \gamma + \delta) & \phantom{-}\frac{1}{2} (\sigma + \gamma +\delta ) \\[1 mm]
\phantom{-} \frac{1}{2} (\sigma + \gamma + \delta) & \phantom{-}\gamma  & - \frac{1}{2} (\sigma+\gamma+\delta )
   & - \frac{1}{2} (\sigma+\gamma+\delta) \\[1 mm]
 \phantom{-} \frac{1}{2} (\sigma-\gamma-\delta) & -\frac{(\sigma+\delta) (\sigma-\gamma-\delta)}{\sigma+\gamma+\delta} & \phantom{-} \frac{1}{2} (\sigma-\gamma-\delta) & - \frac{1}{2} (\sigma+3 \gamma - \delta) \\
\end{array}
\right) \, ,
\eeq
\beq
N_{z,1} \eqsp \left(
\begin{array}{cccc}
 \sigma -\delta  & -\frac{2 (\sigma -\delta ) (\sigma+\delta)}{\sigma+\gamma+\delta}
   & -\sigma+\delta  & \gamma  \\
 \frac{1}{2} (\sigma+\gamma+\delta) & -\sigma-\delta & -\frac{1}{2} (\sigma +\gamma+\delta) & -\frac{1}{2} (\sigma+\gamma+\delta ) \\
 0 & 0 & 0 & \phantom{-\frac{1}{2}} \sigma+\gamma+\delta  \\
 0 & 0 & 0 & -2 \delta  \\
\end{array}
\right) \, ,
\eeq
\beq
N_{z,\bar z} \eqsp \left(
\begin{array}{cccc}
 -4 \, \sigma & 0 & 0 & 0 \\
 0 & 0 & 0 & 0 \\
 0 & 0 & 0 & 0 \\
 0 & 0 & 0 & 0 \\
\end{array}
\right) \, .
\eeq
The result for $N_{\bar z}$ follows exchanging $z \, \leftrightarrow \, \bar z$ and  flipping signs as explained after \eqref{explainSigns}.

Second, we give the equivalent quantities for $\hat I_{2,2}$. The matrices $D_t, \, D_s$ are the only identical objects, all other quantities change:
\beq
C_t \eqsp - \frac{1}{(\beta + \gamma - \delta) \, s^2} \, D_t^{-1} 
\left(\begin{array}{cc}
 -\frac{\alpha- \beta - 2 \gamma }{\alpha-\gamma-\delta} & 1 \\
 1 & -\frac{2 \delta }{\beta+\gamma + \delta} \\
\end{array}
\right) \, D_t^{-1} \, ,
\eeq
\beq
C_s \eqsp -\frac{1}{\beta+\gamma-\delta} \, D_s^{-1} \, \tilde C_s \, D_s^{-1} \, ,
\eeq
\beq
\tilde C_{s} \eqsp \left( \begin{array}{cccc}
 \frac{2 (\beta + \gamma - \delta)}{(\alpha -\gamma -\delta) (\beta +\gamma+ \delta)}
   & 0 & 0 & 0 \\
 0 & \frac{2 \beta  (\alpha -\beta )}{(\alpha -\gamma-\delta) (\beta -\gamma-\delta)
   (\beta -\gamma+\delta)} & \frac{- 2 \alpha  \beta +\beta ^2+2 \beta  \gamma - \gamma^2+\delta
   ^2}{(\alpha -\delta -\gamma ) (\beta -\gamma +\delta) (\beta -\gamma -\delta
   +\gamma )} & \frac{1}{\beta -\gamma -\delta} \\
 0 & \frac{-2 \alpha  \beta +\beta ^2+2 \beta  \gamma -\gamma^2 +\delta ^2}{(\alpha
   -\gamma-\delta) (\beta -\gamma+\delta) (\beta -\gamma - \delta)} & c_{33} & -\frac{\alpha
   +\beta -2 \delta }{(\alpha +\gamma-\delta) (\beta -\gamma-\delta)} \\
 0 & \frac{1}{\beta -\gamma -\delta } & -\frac{\alpha +\beta -2 \delta }{(\alpha +\gamma -\delta) (\beta -\gamma-\delta)} & \frac{2 \delta  (\alpha +\beta - 2 \delta
   )}{(\alpha +\gamma-\delta) (\beta -\gamma-\delta) (\beta +\delta +\gamma )} \\
\end{array}
\right)
\eeq
with
\beq
c_{33} \eqsp \frac{2
   \left(\alpha ^3 \beta+\alpha ^2 \gamma ^2 -\alpha ^2 \delta ^2-\alpha  \beta ^2 \gamma
   -\alpha  \beta  \gamma ^2-\alpha  \beta  \delta ^2+\alpha 
   \gamma ^3-\alpha  \gamma \delta ^2 -\beta ^3 \gamma +\beta  \gamma ^3 +3 \beta  \gamma \delta ^2 -\gamma ^4 +\delta^4 \right)}{(\alpha -\gamma-\delta) (\alpha +\gamma -\delta) (\alpha + \gamma 
   +\delta ) (\beta -\gamma+\delta ) (\beta -\gamma-\delta)} \, .
\eeq
The second level connection becomes
\beq
\Omega_s \eqsp \frac{1}{s} \, \Omega_{s,0} + \frac{1}{s-1} \, \Omega_{s,1} + \frac{1}{s-z} \, \Omega_{s,z} + \frac{1}{s-\bar z} \, \Omega_{s,\bar z} + \frac{1}{s-z \bar z} \, \Omega_{s,z \bar z} \, ,
\eeq
\beq
\Omega_{s,0} \eqsp \left(
\begin{array}{cc}
 \alpha +\gamma + \delta -1 & (1-z) (1-\bar z) (\beta + \gamma +\delta) \\
 0 & \alpha + \gamma -\delta -1 \\
\end{array}
\right) \, , \qquad
\Omega_{s,1} \eqsp
\left(
\begin{array}{cc}
 -\alpha +\beta +2 \gamma  & 0 \\
 \frac{\alpha -\gamma - \delta }{(1-z) (1-\bar z)} & 0 \\
\end{array}
\right) \, ,
\eeq
\beq
\Omega_{s,z} \eqsp \Omega_{s,\bar z} \eqsp \left(
\begin{array}{cc}
 -\beta -\gamma-\delta & -(1-z) (1-\bar z) (\beta +\gamma + \delta) \\
 -\frac{\alpha -\gamma - \delta}{(1-z) (1-\bar z)} & -\alpha + \gamma + \delta  \\
\end{array}
\right) \, , \quad
\Omega_{s,z \bar z} \eqsp \left(
\begin{array}{cc}
 0 & 0 \\
 \frac{\alpha - \gamma - \delta}{(1-z) (1-\bar t)} & \alpha -\beta -2 \gamma  \\
\end{array}
\right) \, .
\eeq


\begin{thebibliography}{99}

\bibitem{BKV}
B.~Basso, S.~Komatsu and P.~Vieira,
[arXiv:1505.06745 [hep-th]].

\bibitem{shotaThiago1}
T.~Fleury and S.~Komatsu,
JHEP \textbf{01} (2017) 130,
[arXiv:1611.05577 [hep-th]].

\bibitem{shotaThiago2}
T.~Fleury and S.~Komatsu,
JHEP \textbf{02} (2018) 177,
[arXiv:1711.05327 [hep-th]].

\bibitem{theMB}
V.~A.~Smirnov,
Springer Tracts Mod. Phys. \textbf{211} (2004). 

\bibitem{usFivePoint}
M.~De Leeuw, B.~Eden, D.~Le Plat, T.~Meier and A.~Sfondrini,
JHEP \textbf{09} (2020) 039,
[arXiv:1912.12231 [hep-th]].

\bibitem{gluingInter}
G.~Crisanti, B.~Eden, M.~Gottwald, P.~Mastrolia and T.~Scherdin,
[arXiv:2411.07330 [hep-th]].

\bibitem{inter1}
K.~Aomoto and M.~Kita,
Springer (2011).

\bibitem{inter2}
P.~Mastrolia and S.~Mizera,
JHEP \textbf{02} (2019) 139,
[arXiv:1810.03818 [hep-th]].

\bibitem{davydychev1}
N.~I.~Usyukina and A.~I.~Davydychev,
Phys. Lett. B \textbf{298} (1993) 363.

\bibitem{davydychev2}
N.~I.~Usyukina and A.~I.~Davydychev,
Phys. Lett. B \textbf{305} (1993) 136.

\bibitem{symbolGonchiLog1}
A.~B.~Goncharov,
Math. Res. Lett. \textbf{5} (1998) 497,
[arXiv:1105.2076 [math.AG]].

\bibitem{symbolGonchiLog2}
A.~B.~Goncharov, M.~Spradlin, C.~Vergu and A.~Volovich,
Phys. Rev. Lett. \textbf{105} (2010) 151605,
[arXiv:1006.5703 [hep-th]].

\bibitem{MBresm}
A.~V.~Smirnov and V.~A.~Smirnov,
Eur. Phys. J. C \textbf{62} (2009) 445,
[arXiv:0901.0386 [hep-ph]].

\bibitem{inter3a}
S.~Mizera,
Ph.D. thesis, Princeton, Institute for Advanced Studies (2020),
[arXiv:1906.02099 [hep-th]].

\bibitem{inter3b}
H.~Frellesvig, F.~Gasparotto, M.~K.~Mandal, P.~Mastrolia, L.~Mattiazzi and S.~Mizera,
Phys. Rev. Lett. \textbf{123} (2019) no.20, 201602,
[arXiv:1907.02000 [hep-th]].

\bibitem{inter3}
H.~Frellesvig, F.~Gasparotto, S.~Laporta, M.~K.~Mandal, P.~Mastrolia, L.~Mattiazzi and S.~Mizera,
JHEP \textbf{03} (2021) 027,
[arXiv:2008.04823 [hep-th]].

\bibitem{howManyMasters}
R.~N.~Lee and A.~A.~Pomeransky,
JHEP \textbf{11} (2013) 165,
[arXiv:1308.6676 [hep-ph]].

\bibitem{hennAlg}
J.~M.~Henn,
Phys. Rev. Lett. \textbf{110} (2013) 251601,
[arXiv:1304.1806 [hep-th]].

\bibitem{ChengWu}
H.~Cheng and T.~T.~Wu,
 MIT Press, Cambridge, MA, USA, (1987).

\bibitem{kosower}
D.~Kosower, https://mbtools.hepforge.org.

\bibitem{conicHulls}
B.~Ananthanarayan, S.~Banik, S.~Friot and S.~Ghosh,
Phys. Rev. Lett. \textbf{127} (2021) no.15, 151601,
[arXiv:2012.15108 [hep-th]].

\bibitem{beukers}
F.~Beukers,
``Notes on A-hypergeometric functions", University of Utrecht.

\bibitem{chinese4}
T.~Feng, C.~Chang, J.~Chen, and H.~Zhang,
Nucl.~Phys. \textbf{B953} (2020), 114952 [arXiv:1912.01726 [hep-th]].

\bibitem{chestnov1}
V.~Chestnov, F.~Gasparotto, M.~K.~Mandal, P.~Mastrolia, S.~J.~Matsubara-Heo, H.~J.~Munch and N.~Takayama,
JHEP \textbf{09} (2022) 187,
[arXiv:2204.12983 [hep-th]].

\bibitem{chestnov2}
V.~Chestnov, S.~J.~Matsubara-Heo, H.~J.~Munch and N.~Takayama,
JHEP \textbf{11} (2023) 202,
[arXiv:2305.01585 [hep-th]].

\bibitem{chestnov3}
G.~Brunello, V.~Chestnov, G.~Crisanti, H.~Frellesvig, M.~K.~Mandal and P.~Mastrolia,
JHEP \textbf{09} (2024) 015,
[arXiv:2401.01897 [hep-th]].

\bibitem{ambre}
J.~Gluza, K.~Kajda and T.~Riemann,
Comput. Phys. Commun. \textbf{177} (2007) 879,
[arXiv:0704.2423 [hep-ph]].

\bibitem{MBm}
M.~Czakon,
Comput. Phys. Commun. \textbf{175} (2006) 559,
[arXiv:hep-ph/0511200 [hep-ph]].

\bibitem{Nakanishi}
N.~Nakanishi,
Gordon and Breach  (1971).

\bibitem{Bluemlein}
J.~Bl\"umlein, I.~Dubovyk, J.~Gluza, M.~Ochman, C.~G.~Raab, T.~Riemann and C.~Schneider,
PoS \textbf{LL2014} (2014) 052,
[arXiv:1407.7832 [hep-ph]].

\bibitem{BDS}
Z.~Bern, L.~J.~Dixon and V.~A.~Smirnov,
Phys. Rev. D \textbf{72} (2005) 085001,
[arXiv:hep-th/0505205 [hep-th]].

\end{thebibliography}
\end{document}